%% file: sn-article.tex
\documentclass[pdflatex,sn-mathphys-num]{sn-jnl}

\usepackage{graphicx}%
\usepackage{multirow}%
\usepackage{amsmath,amssymb,amsfonts}%
\usepackage{amsthm}%
\usepackage{mathrsfs}%
\usepackage[title]{appendix}%
\usepackage{xcolor}%
\usepackage{textcomp}%
\usepackage{manyfoot}%
\usepackage{booktabs}%
\usepackage{algorithm}%
\usepackage{algorithmicx}%
\usepackage{algpseudocode}%
\usepackage{listings}%

\usepackage{lipsum}
\usepackage{caption}
\usepackage{subcaption}
\usepackage{adjustbox}
\usepackage{textgreek}
\usepackage{array}
\newcolumntype{F}[1]{%
    >{\raggedright\arraybackslash\hspace{0pt}}p{#1}}%
\newcolumntype{T}[1]{%
    >{\centering\arraybackslash\hspace{0pt}}p{#1}}%

\hypersetup{urlcolor=blue, colorlinks=true} 

\theoremstyle{thmstyleone}%
\theoremstyle{thmstyletwo}%

\theoremstyle{thmstylethree}%

\begin{document}

\title[Article Title]{Real-Time Integrated Dispatching and Idle Fleet Steering with Deep Reinforcement Learning for A Meal Delivery Platform}


\author*[1]{\fnm{Jingyi} \sur{Cheng}}\email{j.cheng-1@tudelft.nl}

\author[1]{\fnm{Shadi} \sur{Sharif Azadeh}}\email{s.sharifazadeh@tudelft.nl}

\affil*[1]{\orgdiv{Transport and Planning}, \orgname{Delft University of Technology}, \orgaddress{\street{Stevinweg 1}, \city{Delft}, \postcode{2628CN}, \state{Zuid Holland}, \country{Netherlands}}}


\abstract{To achieve high service quality and profitability, meal delivery platforms like Uber Eats and Grubhub must strategically operate their fleets to ensure timely deliveries for current orders while mitigating the consequential impacts of suboptimal decisions that leads to courier understaffing in the future.
This study set out to solve the real-time order dispatching and idle courier steering problems for a meal delivery platform by proposing a reinforcement learning (RL)-based strategic dual-control framework. 
To address the inherent sequential nature of these problems, we model both dispatching and steering as Markov Decision Processes. Trained via a deep reinforcement learning (DRL) framework, we obtain strategic policies by leveraging the explicitly predicted demands as part of the inputs. In our dual-control framework, the dispatching and steering policies are iteratively trained in an integrated manner. These forward-looking policies can be executed in real-time and provide decisions while jointly considering the impacts on local and network levels. To enhance dispatching fairness, we propose convolutional deep Q networks to construct fair courier embeddings. To simultaneously rebalance the supply and demand within the service network, we propose to utilize mean-field approximated supply-demand knowledge to reallocate idle couriers at the local level. 
Utilizing the policies generated by the RL-based strategic dual-control framework, we find the delivery efficiency and fairness of workload distribution among couriers have been improved, and under-supplied conditions have been alleviated within the service network.
Our study sheds light on designing an RL-based framework to enable forward-looking real-time operations for meal delivery platforms and other on-demand services.}




\maketitle

\textbf{This manuscript is currently under review at Data Science for Transportation.}

\section{Introduction}
\label{introduction}
\input{Chapters/Ch1_Introduction}

\section{Literature Review}
\label{chapter:literature}
\input{Chapters/Ch2_Literature}

\section{Problem Description}
\label{chapter:problem_description}
\input{Chapters/Ch3_SystemDescription}

\section{RL-based Strategic Dual-Control Framework}
\label{chapter:methods}
\input{Chapters/Ch4_Methodology}

\section{Model Application}
\label{chapter:instances}
\input{Chapters/Ch5_Instances}

\section{Computational Results}
\label{chapter:result_and_discussion}
\input{Chapters/Ch6_Results}

\section{Discussions}
\label{chapter:discussion}
\input{Chapters/Ch7_Discussion}

\section{Conclusions}
\label{chapter:conclusion}
\input{Chapters/Ch8_Conclusion}

\begin{appendices}
\section{Exploratory Data Analysis of the Full Order Transaction Data}
\label{Appendix_1}
\input{Chapters/Appendix_1}

\section{Reinforcement Learning}
\label{Appendix_2}
\input{Chapters/Appendix_2}

\section{Predictive Performance of the Short-Term Demand Predictor}
\label{Appendix_3}
\input{Chapters/Appendix_3}




\end{appendices}


\bibliography{example}

\end{document}

%% file: Chapters/Ch1_Introduction.tex
The rapid development of information technology has changed the way we live, also the way we `eat'. The On-Demand Meal Delivery (ODMD) service has been one of the fastest-expanding businesses in recent years. The meal delivery platforms connect customers and restaurants by enabling online order placements and providing a convenient click-to-door service to deliver freshly prepared meals from restaurants to households \citep{seghezzi2021demand}. 
During the global Covid-19 pandemic, the market for meal delivery services has more than doubled in the United States \citep{ahuja2021ordering}. 
With an expected annual growth rate of 6.95\% from 2023 to 2027, the on-demand meal delivery has become a 222.50 billion dollar business covering more than 1 billion users around the world \citep{statista}.

From the platform perspective, the interests of customers, restaurants, and couriers should all be taken into account to run a successful business. 
Despite the massive market volume and huge potential in the ODMD industry, the current meal delivery business is highly cost-intensive, as described by the chief operating officer of DoorDash during an interview \citep{doordash}. 
 To establish a solid user base, platforms need to uphold high customer satisfaction, which is closely related to the quality and reliability of service  \citep{koay2022model, banerjee2019measuring}. It means the customers generally have low patience for delay and they expect to receive fresh meals. Reliability relates to providing robust service to deliver the orders correctly and timely according to promise.
In this context, canceling accepted orders due to understaffing of couriers significantly damages customer satisfaction and the trust of restaurants. 
Operation efficiency is crucial for the profitability of ODMD companies. While hiring a larger fleet simplifies the courier-order matching problem, it also incurs higher costs due to employing more couriers than necessary. 
Therefore, platforms need to strategically plan deliveries and optimize courier utilization to maximize profit margins by minimizing operating costs

Attention should be paid to the welfare of couriers in the operational decision process to run a sustainable and ethical ODMD business. In recent years, several courier strikes have occurred in different parts of the world \citep{NewsCornwell2022, NewsTubridy2023, NewsZhao2023}, highlighting two significant concerns from the perspective of couriers. 
The first concern relates to `productivity over safety' due to unreasonable workload. When the system fails to respond to unexpected spikes in demand, couriers may be overwhelmed by the intensive workload and strict performance requirements. The other concern regards the payment.
Currently, most meal delivery service providers operate as gig platforms, which hire couriers as independent contractors with the commission fee per order as their major part of income. But many couriers struggle to earn a satisfying wage. Motivated by this payment scheme, gig couriers prioritize acting in the best interest of their time. It is common for gig couriers to reject orders that are not profitable to them. And they are often overcrowded in popular restaurant areas without communication with each other. The individual decision-making of couriers can lead to service inequalities and lower operating efficiency of the platform. 
The intense competition among couriers also promotes risky riding. 
Recently, the concept of employing hourly-paid couriers has been introduced to protect the welfare of couriers \citep{pulignano2021working}. However, the potential impact of this new operating scheme on delivery performance remains uncertain.  

Operating a successful on-demand meal delivery service is a challenging task not only because of the complex relationship between different parties but also because of its unique characteristics of operation. First of all, the couriers and orders are distributed dynamically within the service network.
Restaurants are typically located in different parts of the city, which differs from other last-mile delivery problems where the pickup location is fixed at the warehouse. Platforms often keep their couriers free-flowing in the city. 
The other aspect to consider is the uncertain preparation time of the order. The actual preparation time often deviates from the expectation. 
To maintain high delivery quality and fleet efficiency, the time of arrival of the courier at the restaurant should ideally be as close as possible to the ready time of order. 
Additionally, since arrivals of orders are not known in advance, the dispatching operations of the current orders should be generated in real-time without perfect information about future orders.
Furthermore, given the perishable property of meals and the customers' low tolerance for delay, meals are delivered the sooner, the better. Therefore, we assume the dispatching decision is made once a new order comes in, instead of waiting for the order buffer to be filled up to a specific size and then generating the decisions altogether. Although bundling is a well-working strategy in same-day package delivery \citep{ma2021hierarchical}, letting a ready-to-go order wait for other similar orders for bundling does not guarantee improvement in service quality nor reduction in operational cost, as shown by \cite{yildiz2019provably}. Hence, couriers are assumed to carry at most one order per trip in our study.

Characterized by the inherent urgency, uncertainty, and dynamic nature of order delivery tasks, as well as the complex multi-objective nature of the business, the on-demand meal delivery service is widely acknowledged as the ultimate challenge in last-mile logistics. It garners significant interest from scholars in both the fields of operations research and machine learning. 

Much attention has been paid to the order assignments of ODMD. Treating the order dispatching problem of ODMD as a combined variant of the classic vehicle routing problems (VRP) and traveling salesman problems (TSP), operations research (OR) methods, such as heuristic and metaheuristic approaches, are proposed to generate optimal bipartite matching between orders and couriers  \citep{reyes2018meal,yildiz2019provably,steever2019dynamic,liao2020multi,gupta2022fairfoody,arslan2022supply}. However, many OR approaches are solved myopically in a rolling window manner without considering the long-term influence of decisions. Additionally, real-time execution is challenging for most OR approaches in a large system. 
To encounter the sequential impact of decisions as well as improve the generation speed of decisions, reinforcement learning (RL)-based approaches are studied to provide real-time decisions according to the learned policy. 
To reduce the computation complexity, studies \cite{jahanshahi2022deep} and \cite{bozanta2022courier} model each courier as a decision agent. But their approaches have come across scalability issues and cannot be applied to more than ten couriers. \cite{zou2022online} introduce a centralized RL-based dispatching framework, which manages to solve large-scale order dispatching problems using historical data collected from a platform. Nevertheless, their model hasn't incorporated the uncertainty in meal preparation times for decision-making since they assume orders are always ready to be picked up upon arrival of the couriers. 

Steering strategies for supply-demand rebalancing aims to proactively match the number of available couriers to the pickup requests from restaurants in the future. While proactive demand steering operations typically relate to dynamic pricing strategy and order rejections, supply steering operations often include the reallocation of idle couriers to the under-supplied area. From the demand steering side, RL-based strategies involving order rejections have been widely discussed in the literature on on-demand meal delivery \citep{jahanshahi2022deep, bozanta2022courier}, as well as other dynamic pickup and delivery services \citep{ulmer2020meso, kavuk2022order}. On the other hand, little attention has been paid to real-time supply steering operations for ODMD, such as idle fleet steering. Splitting the city into several large regions, \cite{wang2023cross} introduce an RL-based framework to dynamically reallocate couriers across regions. Utilizing demand anticipation, \cite{arslan2022supply} introduce an OR optimization framework that dynamically and simultaneously steers the demand and fleet. The possibility of reallocating idle couriers at the micro-level (i.e., from one grid to another) has not been much explored yet in the field of ODMD. In addition, the learning-based framework for order dispatching and steering operations has not been jointly studied in the previous literature of ODMD.

The goal of this research is to design an efficient reinforcement learning (RL)-based strategic dual-control framework that leverages predicted short-term demand information to generate both optimal order dispatching and idle courier steering policies for an on-demand meal delivery platform (ODMD).  Particularly, for this multi-objective on-demand delivery system, we wonder whether the policies learned by our framework can improve the system-level delivery efficiency, ease the under-supplied condition within the service network, and improve the fairness of workload distribution among couriers. We are also curious whether the use of explicitly predicted demand to construct forward-looking policies leads to an improvement in performance. We assume the operational decisions are generated from the platform perspective since we are most interested in whether the meal delivery platform can be a self-organized system. 

The main novelties of our study arise from the following aspects. In order to upscale the limited dataset for the data-intensive training of RL algorithms, we bootstrap the data by creating an order sampler implemented by the parameters estimated from the data. For short-term demand forecasting, a lagged-dependent XGBoost regression model is applied to generate adaptive predictions of demand within the network in real time. Then the predicted demand information is applied to the dual-control framework to assist strategic decision-making of the system.
Making use of a multi-objective reward function, the order dispatching obtained from our RL-based framework aims to balance the delivery performance at both current order and system level. Moreover, our model allows postponing as a possible decision, in case the platform believes a more suitable courier will show up in the near future.
To prevent algorithmic biases, we propose \textit{Convolutional Deep Q Network} (Conv-DQN) as the value function approximator to create fair feature embeddings of couriers through the convolutional filter. To enable decentralized executions while reserving optimally at the network level, our trained idle courier reallocation policy generates local reallocation decisions by using the mean-field approximated supply-demand balancing information of the local neighborhoods. Within this framework, a sequential decision framework is designed to maintain the supply and demand balance within the service network by coordinating the passive and active rebalancing decisions from the order dispatching and fleet steering decisions.
We also introduce the `sandwich learning strategy' to jointly train the order dispatching and fleet steering policies in an iterative manner within our simulated system.

The remainder of this paper is structured as follows. We review the relevant recent literature in Section \ref{chapter:literature}. Then, we describe the on-demand meal delivery system by providing the basic definitions and assumptions in Section \ref{chapter:problem_description}. The structure of our strategic dual-control framework is explained in Section \ref{chapter:methods}, where we introduce the short-term demand prediction model and the design of the RL-based order dispatching and idle courier reallocation methods. 
Next, Section \ref{chapter:instances} elaborates on the dataset, policy training strategy, and performance evaluation scheme utilized in our experiments. The results of our study are presented in Section \ref{chapter:result_and_discussion}, followed by our discussion in Section \ref{chapter:discussion}.
Lastly, we summarize our study, reflect on the limitations of this research, and provide suggestions for future research in Section \ref{chapter:conclusion}.

%% file: Chapters/Ch2_Literature.tex
With the development of technologies, on-demand services have revolutionized the business by offering their users great flexibility in accessing the service anytime they want without the need for prior scheduling. However, along with the benefits, the emergence of on-demand services has also introduced new challenges. On-demand service providers are required to generate decisions in real-time while optimizing the network-level performance with the presence of uncertainty and dynamic arrivals of requests in the system. 
The operational challenges of on-demand meal delivery, along with similar on-demand services like ride-hailing and instant parcel pickup and delivery, can be categorised as subproblems within the broader field of dynamic pickup and delivery problems (DPDP).
In this section, we review the recent studies on ODMD and its closely related problems on the topics of order dispatching and supply steering, with an emphasis in the discussion of incorporation of forward-looking information, consideration of dynamic and stochasticity from the environment, the scalability and real-timeness of solutions.

\subsection{Predict-then-Optimize}
\label{subset:short-term demand}

Demand forecasting has been adopted as the first step in many studies to provide anticipatory insights to enhance the operations of a dynamic system. This sequential method is also known as the `predict-then-optimize' approach, where the parameter for forecasting is firstly estimated to minimize prediction errors, then the predictors are implemented sequentially in the downstream optimization model. \cite{wang2020demand} introduce demand-aware route planning for shared mobility service to improve overall efficiency by considering the spatial distribution of future demand and estimate the likelihood of getting new orders on the route. 
In order to rebalance the supply and demand in the service network, \cite{miao2021data} introduces a vehicle relocation algorithm, where the predicted demand distribution is considered for strategic fleet steering.
For the optimization of order dispatching decisions, \cite{al2019deeppool} propose DeepPool to utilize the predicted demand spatial distribution for a ride-sharing platform.

Among the literature for on-demand delivery, machine learning approaches have been frequently considered for the task of short-term demand forecasting.
\cite{saadi2017demand_ride} employ and compare different forecasting models to predict spatial-temporal demand at aggregated time intervals of 10 minutes for a ride-hailing platform. In their study, the non-parametric machine learning method gradient boosted trees obtains the highest prediction quality among other models. 
In order to predict the per 15-minute demand for taxi service in New York City, \cite{qian2020short_demand_taxi} propose a data-driven approach named boosting Gaussian conditional random field (boosting-GCRF) to provide both robust deterministic and probabilistic predictions. Research that is dedicated to demand forecasting to on-demand meal delivery is still sparse in the literature. Using empirical data from a meal delivery platform, \cite{hess2021real} evaluate the hourly aggregate demand predictions generated by both classic univariate and machine learning predictors. And they conclude that the prediction accuracy of random forest based approaches is higher when only limited data (4-6 weeks) is available. 

\subsection{Order Dispatching}
\label{subset:order dispatching}

The order dispatching problems of the on-demand meal delivery service system are about assigning the orders to the couriers. Various studies have been conducted to investigate the order dispatching problems using OR-based methods. 
\cite{yildiz2019provably} simplify the dispatching task by assuming the perfect order information is known in advance and solving it with a Clairvoyant decision maker. Under the static-deterministic assumption, they find evidence that order bundling can neither improve delivery performance nor reduce operational costs. 
Assuming no autonomous decisions from the couriers, \cite{reyes2018meal} propose a bi-objective algorithm to match couriers and orders optimally within each fixed length time window. By relaxing the deterministic assumption of the system, this study finds that the click-to-door time and meal ready to delivered time only increase mildly if the ready time of orders are uncertain. It is worth noting that the order density of their experiments is rather low. A higher demand level may lead to different conclusions.
\cite{steever2019dynamic} investigate the virtual food court delivery problem where an order from the same customer may contain items from multiple restaurants. They introduce a proactive auction-based heuristic that utilizes anticipation of the system's future state to provide dispatching decisions, which is solved with a mixed-integer linear programming method. 
Next to the operational cost and delivery quality, fairness and other factors have also been included as part of the multi-objective function for dispatching problems.
\cite{gupta2022fairfoody} emphasize long-term fairness among couriers' income as part of the objective in dispatching decisions. \cite{liao2020multi} aim to optimize the order dispatching and routing decisions by maximizing customer satisfaction and fair workload distribution among couriers, as well as minimizing the total carbon emissions for delivery. 

\cite{ulmer2021restaurant} introduce an Approximate Dynamic Programming (ADP)-based method to generate optimal routing and order assignment policies for a meal delivery system under the assumptions of stochastic arrivals and uncertain ready times of orders. 
Modeling each courier in the system as a decision agent, \cite{jahanshahi2022deep} introduce an RL-based framework to train a uniform policy. 
In their simulated environment, orders arrive dynamically to the platform but with exact preparation times.
Positive reward is given when an order is delivered within deadline, while late delivery and order rejection grant penalty. Travelling cost only counts for relocation.
In terms of computational performance, their results only limit to at most 7 couriers with low demand assumptions that might not be applicable to large cities. 
\cite{bozanta2022courier} propose an RL-based approach that models each courier as a decision agent to generate their own routing and assignment decisions. By default, a new order is firstly offered to the available courier nearest to the pickup point. They define a scenario specified reward function such that a large penalty is given if an invalid action is taken. In their results, DDQN achieves higher total rewards. However, they reported at max 3 couriers can be included in experiments due to the scalability issue of their framework.
Modeling the delivery routing and order bunching problems as large but finite MDPs, \cite{zou2022online} introduce a centralized dispatching framework that applies DDQN where the platform dispatches each new order to couriers.   
The state vector contains the restaurant address of the related order and full information about each courier's current route. 
After the dispatching decision is generated, a DP-based model is utilized as the second step to update the related courier's route. A fixed reward is granted to every on-time delivery, while a fixed penalty for each late delivery.
The proposed two-stage approach can effectively manage a large number of couriers within a vast city network. Furthermore, the computation time has shown to be much faster compared to a pure routing planning algorithm in OR. 
However, the meal preparation time is omitted in their design of simulations. And orders are assumed to be always ready upon the arrival of couriers, which might not be the case in practice. 

\subsection{Supply Steering}
\label{subset:supply dispatching}
Fleet steering problems involve efficiently repositioning resources to address current open requests and proactively meeting upcoming orders based on anticipated future demand distributions. For operational convenience, it is a common practice to represent the map using a grid system, interpreting reposition decisions as movements between grids \cite{qin2022reinforcement}. 

Centralized reallocation decision-making has been studied with OR researchers.
The real-time fleet steering problem for ODMD has been studied by \cite{arslan2022supply} with a deterministic clairvoyant model with static order information and a forward-looking probabilistic model incorporating the predictive demand insights and anticipation of uncertainty. The decisions are generated for the whole fleet at each decision point. 

Within the field of RL, most studies have investigated the possibility of decentralized reallocation via generating local decisions from a multi-agent RL framework to match supply and demand distribution. 
Among the literature that adopt multi-agent RL, many studies model each driver or rider as a decision agent \citep{lin2018efficient,shou2020reward,yang2020multiagent}.
To improve system-level coordination, the agent's reward function often needs to be carefully designed to encourage coordination and avoid agents overcrowding at the high-demand areas and motivate relocation to under-supplied areas. For the vehicle steering problem for a ride-sharing system,  \cite{lin2018efficient} propose to split rewards among all the driver agents in the same grid to prevent overcrowding. Dealing with the taxi supply and demand balancing problem, \cite{yang2020multiagent} assign a fixed positive reward to all drivers if the group of relocation actions leads to a supply-demand balance, otherwise assign a minor negative reward to everyone. 
Aiming to solve the cross-regional courier displacement problems to balance supply and demand for ODMD, \cite{wang2023cross} propose a courier displacement framework that provides regional reallocation instruction to couriers. In this multi-agent RL approach, the couriers act individually as homogeneous decision agents that follow the same centralized contextual policy. The relocation action space is limited to the adjacent regions of the courier’s current region.  Each courier's state input contains the shared global state information about the spatial distributions of idle couriers, couriers in service, open orders, and time, as well as the local information about the courier's current grid ID and serving time. At each time step, a reward is received by each courier, which is defined to be the order response expected ratio obtained by the number of responded orders divided by the number of couriers within the same region.

\subsection{Research Gap and Our Contributions}
Realistic assumptions and model designs are crucial to generate solutions with good practical values. The unique stochastic, dynamic, and multi-objective properties, as well as the real-time and forward-looking decision-making requirements make the ODMD operational problems complicated and challenging. Therefore, the solutions proposed for other on-demand applications, such as ride-hailing, cannot be directly applied to ODMD. 

In the practice of ODMD, the ready time of meal is stochastic, unlike the cases where passengers or parcels are ready for pickup at a predetermined location. 
Thus, the order dispatching problem of ODMD needs to decide the optimal timing to send the courier over to avoid wasting the driver's time on waiting while preventing any negative impact on performance due to late arrivals of courier. 
In order to operate under the pressure of high request frequency and strict delivery deadlines, many platforms opt to perform `instant delivery’ where normally only one order is carried per trip \citep{allen2021understanding, cant2019riding}. On the other hand, previous literature concentrates on the order batching strategy by studying ODMD as a variant of dynamic pickup and delivery problem with multiple orders per trip. To fill up these gaps from previous research, we assume the orders to arrival dynamically with stochastic ready time and each courier can carry at most one order at the same time.

Scalability still remains an issue in the order dispatching policy training of \cite{jahanshahi2022deep} and \cite{bozanta2022courier}. Both studies model couriers as the decentralized decision agents in their multi-agent RL frameworks. 
In our study, the convergence of training is sped up by our proposed convolutional DDQN (Conv-DDQN) algorithm, where fair embeddings of couriers are first created to reduce the information dimension. Additionally, the state vector in our approach is more concise compared to those from previous studies. 
Making use of the predicted demand in the near future, the multi-objective reward function for order dispatching jointly considers the future supply and demand distribution, the sunk cost of the courier’s waiting time at the restaurant, and the delivery efficiency of the current order altogether. By reactively rebalancing supply and demand while reserving delivery efficiency, the dispatching decisions can achieve system-level coordination without direct communication between orders.  

Although fleet steering problems have been extensively studied for taxis and ride-hailing systems, little attention has been paid to real-time steering for meal delivery couriers. Previous research that applies RL for meal delivery problems often model couriers as decision agents who relocate themselves under the motivation of personal incomes \citep{jahanshahi2022deep, bozanta2022courier,wang2020demand}. 
It has not been explored that whether the platform can provide reallocation instruction to its couriers to directly rebalance the supply and demand within the service network. In this study, we study a real-time idle courier reallocation policy that aims to resolve the spatial supply and demand distribution mismatches.

To date, the discussion about designing a `predict-then-optimize' framework with RL-based techniques for ODMD operation problems is still sparse in the literature. Therefore, our study aims to fill the gap by investigating the potential advantages of leveraging demand predictions to develop strategic order dispatching and idle fleet steering policies. We compare their performance against that of myopic policies, which are trained without anticipatory demand information.

For performance evaluation, existing literature on meal delivery planning problems adopts the average order response rate and rate of before-deadline delivery as the primary performance measures for service quality. However, few studies have analyzed operation performance measures, such as courier utilization efficiency and potential assignment bias among couriers. In this study, we propose additional metrics to comprehensively evaluate the balance between supply and demand in the network, pickup distances from couriers to restaurants, and the fairness reflected in the workload distribution among couriers.

%% file: Chapters/Ch3_SystemDescription.tex
In this section, we provide an overview of the on-demand meal delivery platform.
To begin with, we describe how an on-demand meal delivery platform connects different parties during the delivery process in Section \ref{subsect:system}. Then, we explain the fundamental concepts of a meal delivery system and introduce the modeling assumptions we have made in this study. 

\subsection{On-Demand Meal Delivery System}
\label{subsect:system}
From placement to delivery, the process of order involves four parties: the platform, restaurant, courier, and customer. 
When a new order is placed by the customer on the platform, it notifies the platform of the order details. Next, the order is forwarded to the associate restaurant, which sends a confirmation back to the platform, providing an expected meal preparation time. While the meal is being prepared in the restaurant's kitchen, the platform is responsible for assigning the order to a working courier in the service area. 
The process of order and courier matching is known as \textit{order dispatching}. 
Following the platform's dispatching instructions, the courier heads to the restaurant to pick up the meal when he/she is available. The courier collects the order and heads out to the customer if the meal has been prepared. Otherwise, the courier should wait until the meal is ready for delivery. The delivery is completed when the order is handed to the customer at the doorstep. After completing all assigned tasks, we assume the couriers to remain idle at the destination area of their last active task, awaiting further instruction from the platform. To enhance delivery efficiency, the platform proactively reallocates idle couriers towards the area where (is expected to) require more couriers for order pick-up. Throughout operations, the platform collects information on the spatial distribution of idle couriers and anticipates the demand distribution in the city. To rebalance the future supply-demand spatial distribution, the platform sends reallocation instructions to couriers who have been idle for some time. 
The decision-making process for empty fleet reallocations is often referred to as \textit{idle courier steering}. 

\begin{figure}[h!]
    \centering
    \includegraphics[width= 0.8 \columnwidth]{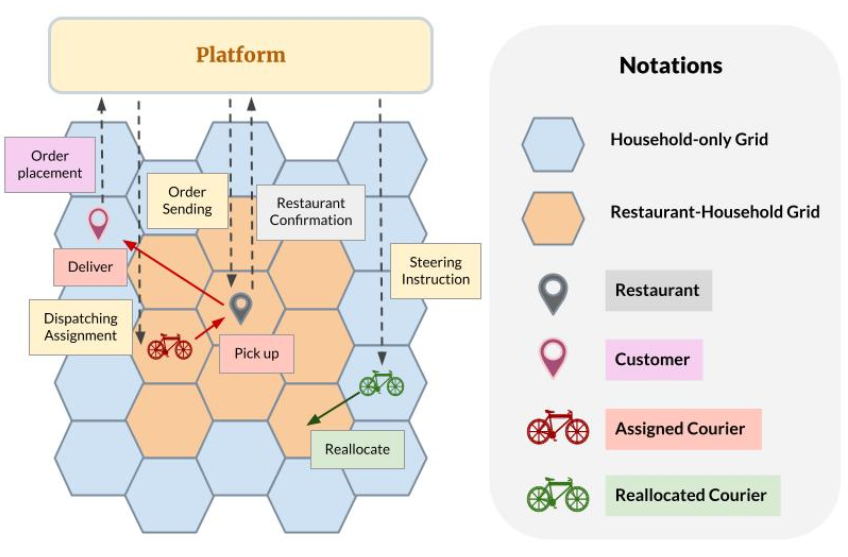}
    \caption{Example processes of order dispatching and idle courier steering.}
    \label{fig:example_process}
\end{figure}

Figure \ref{fig:example_process} visualizes the process of order dispatching and idle courier steering, illustrating the connections between platform, restaurant, courier, and customer for a given example. For management convenience, it is a common industrial practice to discretize the service area into hexagonal grids, with each grid as a unit for operation planning. For each order, the grid where the restaurant is located is called the restaurant grid. And the grid where the household is located is called the household grid. Based on the components, grids can be categorized into two types, household-only grids, and restaurant-household grids.

\subsection{Grids, Distances and Travelling Speed}
To protect user privacy, the original addresses are hashed into hexagonal grids according to Uber's H3 spatial indexing system at resolution level 8 in our empirical dataset \citep{uberh3}. 
Each hexagonal grid on the map is sized identically, covering an area of 0.74 $\text{km}^2$ with an edge length of about 0.53 $\text{km}$. 
Hence, the distance is around 0.92 $\text{km}$ to travel from the center of a grid to the center of its adjacent grid. We assume the biking speed of couriers to be about 16-17 $\text{km/h}$ on E-bikes. 

\begin{figure}[h]
     \centering
     \begin{subfigure}[b]{0.45\textwidth}
         \centering
         \includegraphics[width= \textwidth]{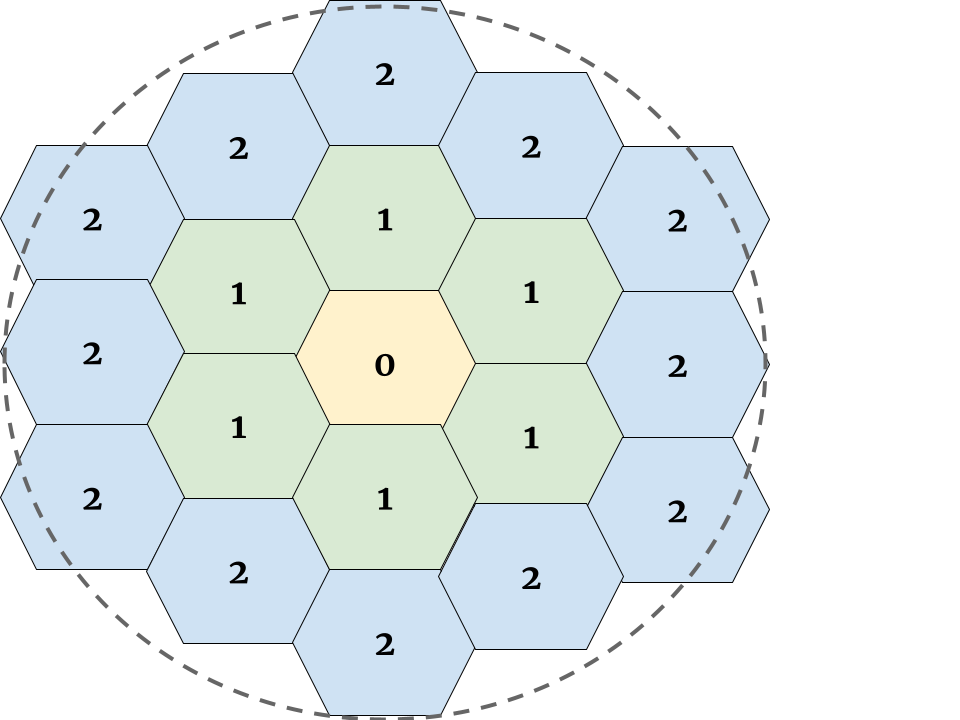}
         \caption{When traveling distance within the same grid (colored in yellow) is 0 unit; traveling distance to adjacent grids (colored in green) is 1 unit; travelling distance to grids two layers away is 2 units, etc.}
         \label{fig:example_distance}
     \end{subfigure}
     \hfill
     \begin{subfigure}[b]{0.45\textwidth}
         \centering
         \includegraphics[width=\textwidth]{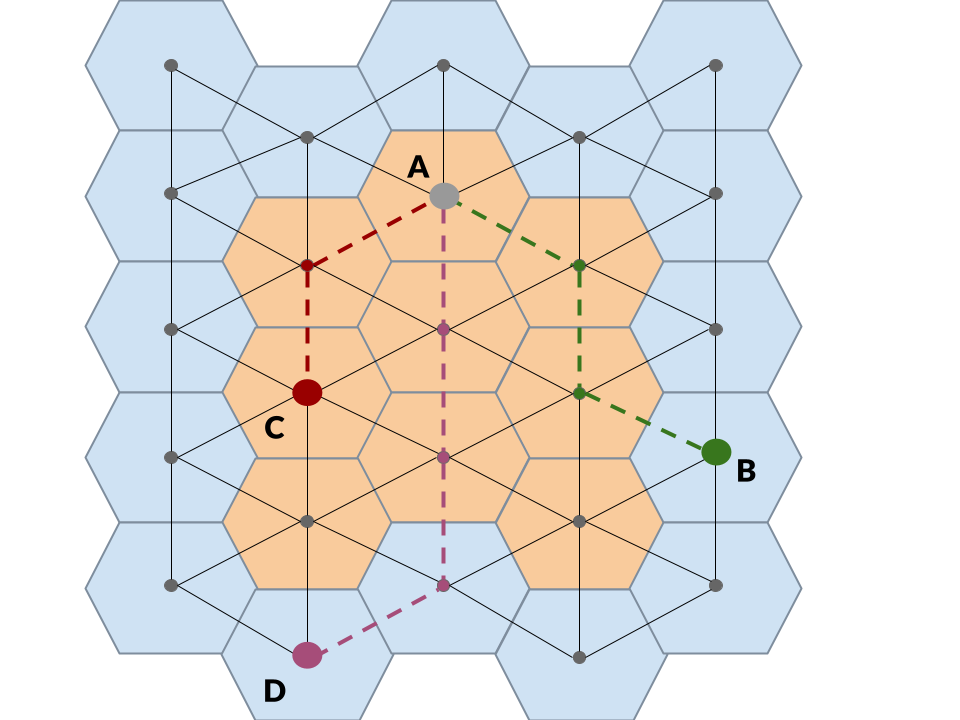}
         \caption{Examples of shortest grid-wise distance calculation in our sample network. The travelling distance from grid A to grid B is 3 units, to grid C is 2 units, to grid D is 4 units. The colored paths are example shortest routes.}
         \label{fig:distance_in_network}
     \end{subfigure}
\caption{Visualized examples for grid-wise distance calculation.}
\label{fig:grid_distance_map}
\end{figure}

Due to the unavailability of specific addresses, we are unable to provide detailed route planning between locations. To simplify the problem, we assume the traveling time to be constant between a given pair of grids, irrespective of the specific locations within those grids. Also, we define the travel distance between the centers of adjacent grids to be one unit, and the travel time speed of couriers to be 3 minutes per unit of distance.
The travel distance is determined based on the number of grid layers between the grids. As illustrated by the example in Figure \ref{fig:example_distance}, the travel distances are determined as follows: it takes zero units of distance to travel within the same grid, one unit to adjacent grids on the first surrounding layer, and two units to the grids at the next layer, and so on.
A network can be created by connecting the hexagonal grid centers in a service region. Based on our assumptions, this network forms a triangulated structure with equal-length edges, making it easy to determine the shortest routes between grids. In Figure \ref{fig:distance_in_network}, we present several examples of shortest paths between grids. It's important to note that there exist multiple alternative shortest paths for travelling between non-adjacent grids in the network.

\subsection{Couriers}

In this research, the couriers are assumed to be employed by shifts. They are paid on an hourly basis with a small amount of commission fee for each delivery or reallocation task they perform. But the hourly payment serves as the primary source of income for the couriers.
For the same shift, we assume the fleet is planned ahead with a fixed number of couriers. During a shift, the scheduled couriers stay active for task assignments. They cannot sign in late to work or sign out early from work as they wish.  

Based on the task being executed, a courier can have one of the five different statuses. If the courier is not currently assigned to any tasks, the courier is idle. Otherwise, he/she is busy. During a delivery task, the courier's status will be `on the way to pick up’ if they are en route to the restaurant, `waiting at the restaurant' if they have arrived but the meal is not yet prepared, and `on the way to delivery' if they have collected the order and are heading to the customer. When the courier is reallocated to an adjacent grid, the status will be `reallocating'. Additionally, each courier maintains an individual queue of his/her on-going tasks. We restrict the couriers from holding more than two delivery tasks simultaneously. It implies that a courier who is already busy with delivery may be assigned at most one additional order. However, couriers who are currently reallocating can be assigned two delivery tasks. Moreover, the reallocation task can only be assigned to couriers who have been idle for more than five minutes.

In our experiment, we consider a fleet of 25 active couriers for the two-hour shift we simulate. At the start of the shift, the couriers are idle, indicating they have no ongoing tasks or preassigned deliveries. Also, their initial locations are randomly chosen from the grids within the sample service region.

\subsection{Orders and Order Sampler}
\label{subset:order}
Orders are received dynamically in the system, which means that the platforms do not possess perfect future knowledge to schedule trips in advance. Additionally, the preparation time for meals is often uncertain, especially during busy dinner hours when the kitchen is catering to both in-house customers and delivery orders. 

Having observed that the demand level varies among different restaurant-household grids in Section \ref{subsect:data}, we decide to sample the dynamic arrivals of orders for each restaurant-household grid $g^r \in G^r$ independently, following an inhomogeneous Poisson process with time-dependent arrival rates of orders $\lambda(t)$. And we estimate the hourly order arrival rates $\lambda_{g^r}(h)$ of restaurant-household grid $g^r$ by the average number of orders received during hour $h, h\in\{19,20\}$ from historical data. The simulation time step of this study is set to be 1 minute. Therefore, the expected total number of orders sampled per minute at hour $h$ is $\sum_{g^r \in G^r} \frac{1}{60} \lambda_{g^r}(h)$. 

Given the restaurant grid of an order is $g_r$, the household grid $g_h$ of this order is chosen randomly based on the pairwise origin-destination probabilities $\mathbb{P}(g^r,g^h), g^h \in G^h$. These probabilities are the estimated likelihoods from the historical transaction data for an order to be picked up from $g^r$ and delivered to $g^h$. Hence, these pairwise probabilities fulfil the relation $\sum_{g_h \in G_h} \mathbb{P}(g^r,g^h) = 1$.
The \textit{estimated preparation duration} $\hat{\tau}_o$ (measured in minutes) of order $o$ is randomly sampled from a normal distribution with a mean of 10 minutes and a variance of 2:
$\hat{\tau}_o \sim \mathcal{N}(\mu=10,\,\sigma^{2}=2)$. The estimated preparation time is shared by the restaurant with the platform upon confirmation of the new order, which is available for decision-making. 
Additionally, we sample the \textit{actual preparation duration} $t_o$ of order $o$ as the estimated preparation time with a small deviation factor drawn from a standard normal distribution: $\tau_o = \hat{\tau}_o + \epsilon$, where $\epsilon \sim \mathcal{N}(\mu=0,\,\sigma^{2}=1)$.

%% file: Chapters/Ch4_Methodology.tex
In this section, we propose the reinforcement learning-based dual-control framework with both strategic order dispatching and idle courier reallocation implemented. In Section \ref{subsect:pred}, we introduce the short-term demand forecasting method which provides the anticipated demand distribution that we later use to generate strategic insights into the service network. 
Then, we formulate the order dispatching problem and illustrate our design of an RL-based model for dispatching decision-making in Section \ref{subsect:order_dispatching}, followed by our problem formulation and model description for idle courier reallocation in Section \ref{subsect:fleet_steering}. Lastly, we explain how the dispatching and steering decisions are made sequentially in our dual-control framework in Section \ref{subsect:dual-control}. In Appendix.\ref{Appendix_2}, we first provide a recap of reinforcement learning techniques. Then, we describe the \textit{Double Deep Q Networks} (DDQN) algorithm and related hyperparameters we utilize for dispatching and reallocation models.

\subsection{Short-Term Demand Prediction with XGBoost}
\label{subsect:pred}

By integrating short-term predictive insights into the system, the platform can generate proactive decisions in real-time. Recognizing the dynamic demand patterns of the meal delivery system and the need for real-time decision-making, we employ an adaptive forecasting method that enables fast-generating demand predictions for the service network. To ensure operational flexibility and minimize noises in prediction, we have opted for a short-term prediction window of 15 minutes. 

In the field of forecasting, the classic parametric time series models, such as exponential smoothing and Autoregressive Integrated Moving Average (ARIMA), are univariate models, where only the historical values of the target variable are used as input features. They are good at interpreting the univariate sequential relations between the previous values and the target of prediction. And the predictions are adaptive to the recent changes in the dynamics of demand. However, these forecasting models require expertise and manual efforts to select a proper formulation. Nevertheless, the forecasting is limited to the recent observations of demand as input, which makes the identification of patterns implicit and potentially challenging. 
Temporal features like the day of the week and the hour of the day can be used to describe the pattern of demand. Known for the capability of capturing the non-linear relations between features and the target variables to predict demand patterns, machine learning (ML) techniques such as random forest regression start to gain popularity in short-term demand predictions. Classic time series forecasting methods often require professional knowledge in selecting the components before parameter estimation. Although this manual selection is not required in ML methods, ML predictors normally require a sufficient amount of data to uncover the demand patterns.

Joining the strengths of both approaches, we employ the lagged-dependent eXtreme Gradient Boosting (LD-XGBoost) as our short-term demand forecasting algorithm to predict the number of orders expected in the next 15 minutes for each grid with restaurants in the service network. 
Introduced by \cite{chen2016xgboost}, XGBoost is an advanced non-parametric ensemble method in machine learning, derived from random forest. It excels in fast training and utilizes decision trees to classify data into scenarios based on features. It is robust against overfitting due to regularization terms and the ensemble nature of the algorithm.
In our model, we include the temporal features, day of the week, and hour of the day, as the inputs of LD-XGBoost. 
Motivated by the usage of the recurrent target variable in autoregressive analysis and Long short-term memory networks (LSTMs) for forecasting, LD-XGBoost also includes four lagged-dependent features from previous time windows as additional inputs. The lagged-dependent inputs are denoted by $y_{g_s}(t,t-15)$, $y_{g_s}(t-15,t-30)$, $y_{g_s}(t-30,t-45)$,$y_{g_s}(t-45,t-60)$, where $y_{g_s}(t_1,t_2)$ indicates the number of orders received by grid $g_s$ between the time period from $t_2$ to $t_1$. Hence, the LD-XGBoost captures the orders received from the previous four 15-minute time windows, covering the demand observed from the previous hour.

\subsection{Strategic Order Dispatching}
\label{subsect:order_dispatching}

The sequential decision-making of the order dispatching problem for ODMD can be formulated as a Markov Decision Process (MDP), since the next dispatching decision inline for the following order can be considered only related to the current decision and its consequence in state changes.
Considering the platform itself as the \textit{decision agent}, we design a RL-based method to find the optimal policy to automatically dispatch the orders that arrive dynamically to the platform with uncertain information. 

\paragraph{\textbf{State}}
The environment state vector is defined based on the urgency of the order, the courier's occupancy information, distance to the restaurant grid, and the supply-demand condition of their last active grid. Given an order $o$, the state vector $s^t_{o} = [\Delta t^r_o, X^t]$ contains the expected remaining ready time of the order $\Delta t^r_o$, and the couriers information set $X^t_C$. The couriers information set is composed by each courier's specific features $X^t_c = (\Delta t_c, d(g^c,g^r_o),\widehat{SD_{g^c,t}}) \in X^t_C$, where $\Delta t_c$ is the expected remaining time to be idle, $d(g^c,g^r_o)$ is the pickup distance between courier and restaurant grids, and $ \widehat{SD_{g^c,t}} = \widehat{N^C_{g^c,t}} - \widehat{N^O_{g^c,t}}$ being the anticipated supply-demand gap of the (future) idle grid for each courier $c$ in the group of couriers $C$. Note that $g^c$ is the courier's (future) idle grid, $g^r$ is the restaurant grid of the order, $ \widehat{N^C_{g^c,t}}$ are the number of idle couriers who are arriving at $g^c$ in the next 15 minutes duration, and $\widehat{N^O_{g^c,t}}$ is the predicted number of orders receiving by $g^c$ in the next 15 minutes period. We call this state vector \textit{strategic} since it utilizes forward-looking information of the network in the future. On the flip side, the state vector is \textit{myopic} if we replace the predicted supply-demand gap elements in the state vector with the current supply-demand gap information of the grids, i.e., 
$ SD_{g^c,t} = N^C_{g^c,t} - N^O_{g^c,t}$.

\paragraph{\textbf{Action}}

In this study, we assume the platform should attend to the orders as soon as they arrive in the system. Next to assigning the order to a courier, the platform can also choose to postpone the dispatching of this order and dispatch it later to receive a higher expected reward. When the platform opts to postpone an order, it receives a penalty for the risk of adding operating pressure to future assignments. And the order will be put back into the order buffer and wait for another dispatching decision in the next minute.
Therefore, there are $|C|+1$ candidate actions in the system action space $\mathcal{A}$ for dispatching decisions. $|C|$ is the number of active couriers in the system, equalling to the fleet size. 
An action filter is implemented to constraint the dispatching decision to select a courier with less than two ongoing delivery tasks or postpone the order. Given that we follow $\varepsilon$-greedy for action selection, the action filter vector simply masks the estimated Q values to be huge negative numbers for the invalid actions.  

\paragraph{\textbf{Reward}}
The dispatching for each order is the joint decision made considering the performance trade-off at the current order and network levels. 
At the current order service level, the customer satisfaction level is positively related to the delivery speed. Customers prefer their orders to be served as soon as it is freshly prepared at the restaurant. That means the time gap between meal ready and courier arrival time, as defined in Equation \eqref{eq:time_gap}, should be preferably close to zero or negative. A positive time gap will be significantly punished for diminishing customer experience. 
At the network level, the operation cost paid for dispatching should be the lower, the better. In our assumptions, the number of couriers scheduled for a shift is fixed. And the couriers are paid per hour by the platform. 
To improve delivery efficiency, the distance from the selected courier to the restaurant for order pickup should be minimized to reduce the time spent on travel. Furthermore, the waiting time of couriers at the restaurant should also be as little as possible. If the courier arrives way before the meal is prepared, much time will be wasted on waiting. In this case, although the delivery time of this current order is minimal, the waiting time of the courier becomes a sunk cost to the platform. It may be prevented by assigning a courier to arrive just before the order becomes ready. Thus, both waiting time and pickup distance are penalized in the reward function to enhance network-level delivery efficiency.
Higher delivery efficiency in the future is possible by strategically considering the supply-demand balancing within the network. In our reward design, we encourage the consideration of `passive rebalancing' in dispatching decision-making by awarding selecting a courier from an over-supplied grid and giving a penalty for choosing a courier from a currently under-supplied grid.
Nevertheless, the cancellation of overdue orders due to the failure of the platform's operation should be heavily penalized. In reality, customers or restaurants may lose their patience and request order refunds if there is an excessive delay in assigning a courier, resulting in a substantial negative impact on the platform's reputation and customer experience.

Considering these four different aspects of performance, the design of our multi-objective reward function for order dispatching is given as the following: When the dispatching action is to assign the order $o$ to courier $c$, the reward is defined by
\begin{equation} \label{eq:dispatching_reward}
    r(s^t_o,c) = 100 + \rho_1 \cdot max \{t^c_o - t^r_o,0\} + \rho_2 \cdot max\{ -(t^c_o - t^r_o) ,0\} + \rho_3 \cdot d(g^c,g^r_o) + \rho_4 \cdot \mathcal{C}(\widehat{SD_{g^c,t}}>0),
\end{equation}
where the basic reward for delivery of an order is 100, $\rho_1, \rho_2, \rho_3, \rho_4$ are the penalty/reward parameters for positive time gap, negative time gap, travel distance for pickup and passive rebalancing, respectively.
Binary variable $\mathcal{C}(\widehat{SD_{g^c,t}}<0)$ check whether the anticipated supply-demand gap is smaller or larger than zero,
\begin{equation} \label{eq:gap_condition}
   \mathcal{C}(\widehat{SD_{g^c,t}}<0) = \begin{cases}
                                            +1,& \text{if } \widehat{SD_{g^c,t}}>0;\\
                                            -1,& \text{otherwise}.
                                        \end{cases}
\end{equation}
In this study, we select the penalty/reward parameter values: $\rho_1 = -5$ per unit of time, $\rho_2=-1$ per unit of time, $\rho_3=-3$ per unit of distance, $\rho_4 = 5$ as a constant.
If the selected action is postponing the order (i.e., $a=p$), an immediate penalty $r(s^t_o,p)=-10$ is provided. However, if the actual ready time of the order is lower than -10, meaning the order failed to be assigned within 10 minutes since it has been prepared, a huge penalty $r(s^t_o,p)=-100$ is granted. And instead of being postponed to the next time step, the order is removed from the system.

\paragraph{\textbf{State Transition}}

When dispatching action has been decided for the current order, the environment updates the system's couriers and order information accordingly. There are two types of updates depending on the selected action. 
When the order has been dispatched or removed due to overdue, the next state vector $s'^{t}_{o'}$ is generated for the next order in line if there are orders still waiting for a decision in the buffer. If the current order is the last one in the queue, a dummy next state $s'^{t}_{o}$ is created as if the next order has the same expected remaining ready time as the current order $o$, but with all couriers' timer ticked forward by one time step in this next state. On the other hand, if the order is postponed to the next time step, the next state $s'^{t}_{o}$ is defined around the same order, and the distribution of courier within the network is assumed to remain unchanged, but with both of the order and couriers' timer ticked forward by one time step.

\begin{figure}[h]
    \centering
    \includegraphics[width= \textwidth]{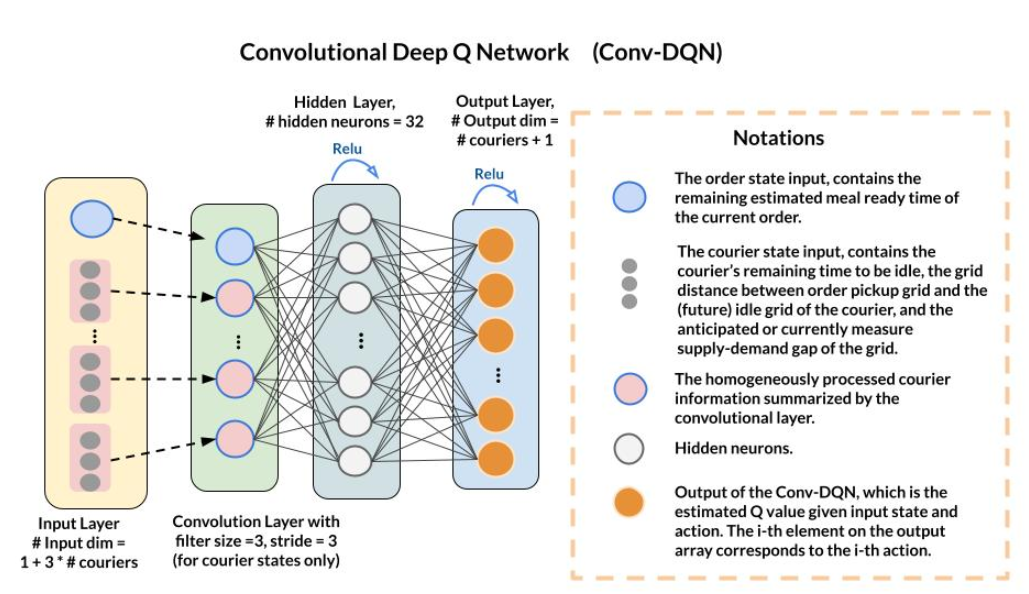}
    \caption{Convolutional Deep Q Networks}
    \label{fig:conv_ddqn}
\end{figure}

\paragraph {\textbf{Algorithm: Convolutional Deep Q Networks}}
For the implementation of the value-based RL framework for order dispatching, we adopt the DDQN algorithm proposed by \cite{van2016deep} (see Appendix.\ref{Appendix_2} and Algorithm \ref{alg:DDQN} for a further explanation).
However, in order to simplify and accelerate the training of the Q value estimations, we propose the \textit{convolutional deep Q network} (Conv-DDQN) as the value estimator for the order dispatching algorithm to replace the simple all-connected Q network. Distinguished from an ordinary all-connected neural network, we use the assumption of homogeneous couriers and create fair embeddings of the state input values related to each courier via a convolutional layer of window size and stride size equalling 3. Illustrated in Figure \ref{fig:conv_ddqn}, embedding of each courier is generated by, 
\begin{equation}
    e_c = \beta_1 \Delta t_c + \beta_2 d(g^c,g^r_o) + \beta_3 \widehat{SD_{g^c,t}},
\end{equation}
with the convolution window parameters $\beta_1, \beta_2, \beta_3$ estimated together with the other parameters in the value estimation neural network. The obtained courier embeddings are then processed together with the order feature $\Delta t^r_o$ through a hidden layer with 32 neurons and Rectified Linear Unit (ReLU) activation. Lastly, the Q value estimates for the actions are generated by the output layer with ReLU activation.
There are two advantages of Conv-DDQN. First, the number of parameters is significantly reduced on the first process layer. So the computational cost for training is reduced. Second, the fair embeddings ensure the couriers' information is processed homogeneously and prevent estimation biases caused by the element position difference in the input vector.

\subsection{Strategic Idle Fleet Steering}
\label{subsect:fleet_steering}

\begin{figure}[h]
    \centering
    \includegraphics[width= 0.85 \textwidth]{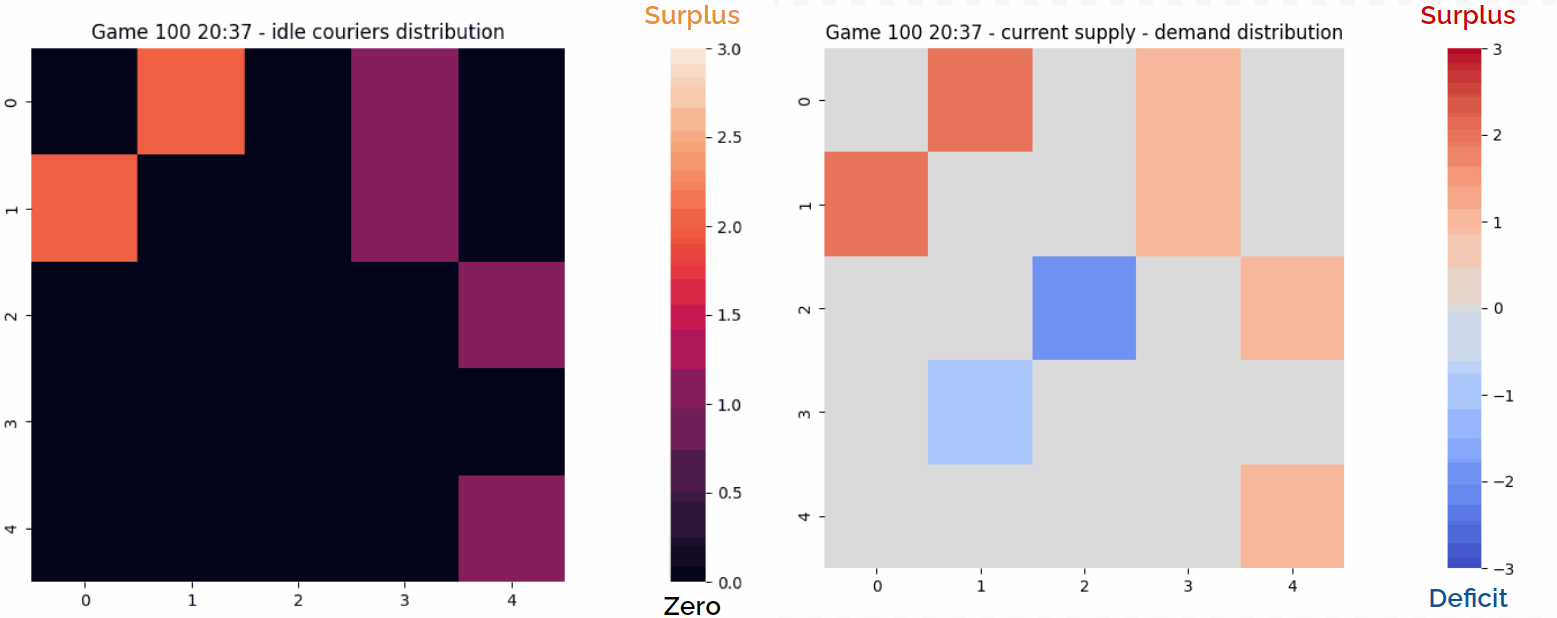}
    \caption{Snapshots of the meal delivery service network at 20:37 from an arbitrary simulation. The figure on the left shows the distribution of idle couriers, while the figure on the right shows the distribution of the current supply-demand gap within the network. The demands arose from the center area, but the idle couriers were located outside the center.}
    \label{fig:no_steering_example}
\end{figure}

If only dispatching instructions are given within the service network, the couriers will stay idle at the household location of the last active delivery task since we assume the salary of couriers is paid per hour and expect no autonomous behaviors from the couriers. 
And the couriers only react to the delivery instructions and start moving upon the arrival of the next delivery task. In Figure \ref{fig:no_steering_example}, we visualize the distribution of the supply-demand gaps and the distribution of idle couriers within our service network when only strategic order dispatching policy is implemented. This example snapshot of the network shows an obvious mismatch of the open orders and courier resources. The idle couriers waiting outside the demand center could have moved towards the restaurant area, which is likely to increase their chance of being matched for future delivery tasks sooner and save time on traveling to the next pickup location. 
To proactively rebalance the supply-demand distribution, we introduce an idle courier steering policy to reallocate the couriers in real time, next to the dispatching instructions. And the steering decisions aim to optimize the future delivery efficiency at the network level considering the trade-off to temporally putting the idle couriers on task.

For each time step, the platform collects the couriers who have been idle for more than five minutes. For each qualified idle courier, we consider the grid that holds the courier as the \textit{decision agent}, which decides where the courier should be reallocated based on the local information. We model the reallocation decision-making process as an MDP and adopt the DDQN algorithm to train an optimal idle courier reallocation policy that can be decentralizedly applied at the local grid level across the service network. The structure of the deep Q network consists of two hidden layers, with 32 and 16 neurons, respectively. Both hidden layers apply ReLU activation.

\paragraph{\textbf{Action}} We assume the reallocation only happens within the immediate neighborhood of the idle courier. It means the courier can only be reallocated to an adjacent grid or stay at the current grid.
Hence, the action set for the reallocation decision corresponds to the valid grids in the immediate neighborhood of the current grid $g^c$ of idle courier $c$.
The number of actions for idle couriers on the edge of the service region is less than seven since the courier cannot be reallocated outside of the region. In this case, an action filter is implemented to avoid invalid action selections.

\begin{figure}[h]
    \centering
    \includegraphics[width= 0.75 \textwidth]{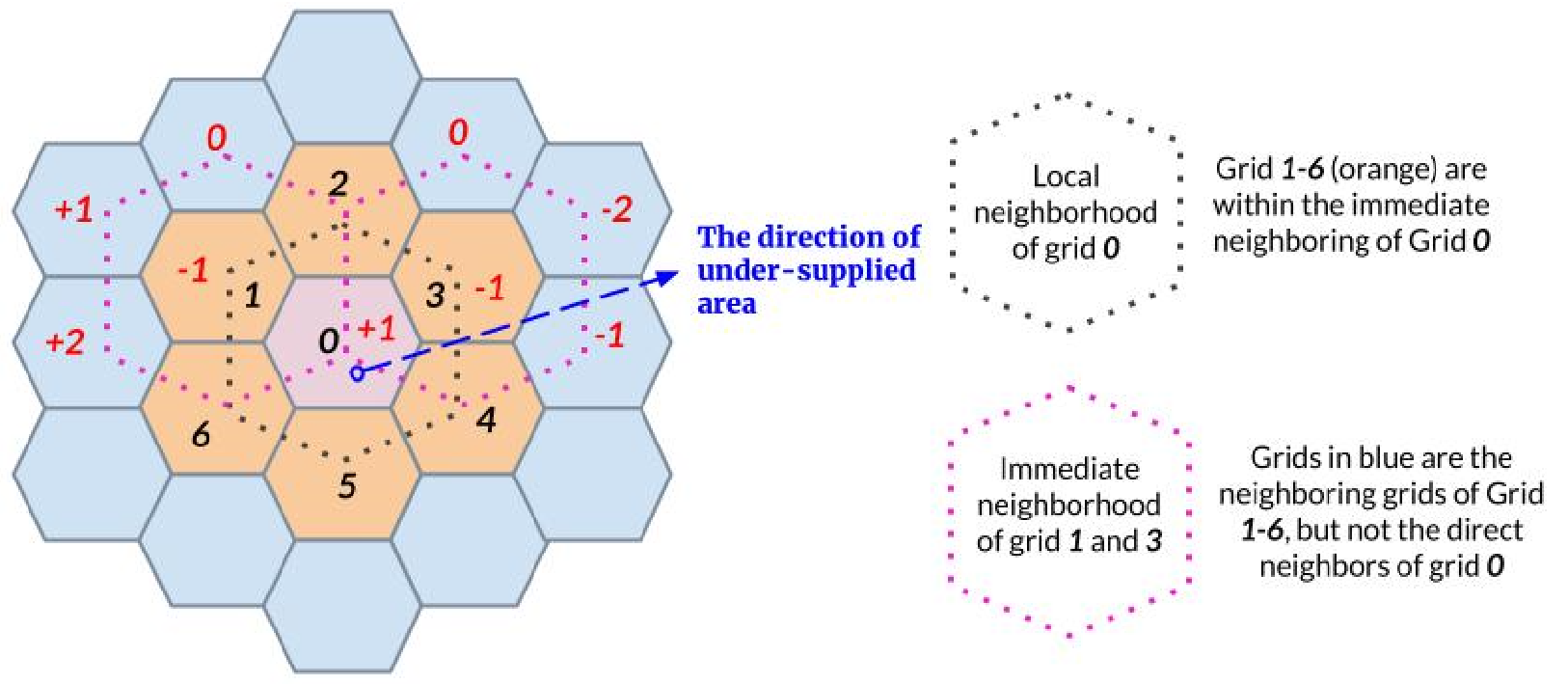}
    \caption{Efficient idle courier steering should be directed towards under-supplied areas. In this figure, the red numbers denote the anticipated or current supply-demand gap on the corresponding grids. And the black numbers denote the ID of the grids within the immediate neighborhood of \textit{grid 0}. The direction towards \textit{grid 3} points to the under-supplied area according to the neighborhood supply-demand information of \textit{grid 3}.}
    \label{fig:idle_steering_direction}
\end{figure}

\paragraph{\textbf{State and State Transition}}
Ideally, the reallocation should steer the idle courier toward the direction of the under-supplied area. In the example shown in Figure \ref{fig:idle_steering_direction}, \textit{grid 0} has an idle courier, while \textit{grids 1} and \textit{3} both have a supply-demand gap of -1. If we only consider the grid-wise supply-demand information of the immediate neighborhood of \textit{grid 0}, \textit{grids 1} and \textit{3} will be regarded as equally good locations to reallocate the courier since they are both under-supplied by one courier. However, this type of short-sighted reallocation can be sub-optimal.
If we also consider the supply-demand level within the immediate neighborhoods of the candidate grids for reallocation. It is simple to tell that the direction pointing at \textit{grid 3} is the direction to the under-supplied area. Thus, the reallocation is likely to be more efficient if we reallocate the courier to \textit{grid 3} instead of \textit{grid 1}. Incorporating further-sighted information helps to generate more efficient reallocations in the network.

\textit{Mean-field approximation} is a trick to approximate the effect of a group of connected individuals by replacing the detailed interactions among them by taking the average effects among individuals. Inspired by this knowledge from the field of complex systems, we summarize the neighborhood supply-demand balancing situations of each candidate grid $g'$ by its local supply-demand score $\textit{score}^t_{g'}$ at time $t$. The grid-wise local supply-demand balance score is simply the sum of anticipated supply-demand gaps in the immediate neighborhood, which is defined as
\begin{equation} \label{eq:mfsd_score}
    \textit{score}^{t}_{g'} = \sum_{g'' \in G_N(g')} \widehat{SD_{g'',t}},
\end{equation}
where $g'$ is a grid from the immediate neighborhood $G_N(g^c)$ of the idle courier's current grid $g^c$ (i.e., $g' \in G_N(g^c)$), and $\widehat{SD_{g',t}}$ is the anticipated supply-demand gap of $g'$. 

The local state vector $s^t_{c}$ specific for idle courier $c$ contains the anticipated supply-demand gaps of all the grid within the immediate neighborhood of its current grid and their grid-wise local supply-demand scores.
Again, if the framework adopts myopic information only, the anticipated supply-demand gaps are replaced by the current supply-demand gaps for both state and reward generations.
The next state $s'^t_{c}$ is obtained by updating the system’s supply and demand distributions after reallocation.

\paragraph{\textbf{Reward}}
We incentivize movement from over-supplied to under-supplied areas and penalize reallocation in the opposite direction.
If the decision is to stay at the same grid, the reward is 0 since no reallocation happens. If the decision is to reallocate the courier from grid $g^c$ to grid $g'^c$, the reward consists of three components. The first and second components $r_1$ and $r_2$ correspond to the anticipated supply-demand gap for the original grid $g^c$ and grid $g'^c$ the courier is reallocated to, respectively. 
The third component $r_3$ is the average local balance score improvement from all the immediate neighboring grids $g' \in G_N(g^c)$, 
\begin{equation*}
    r_3 = \frac{1}{|G_N(g^c)|} \sum_{g' \in G_N(g^c)} \textit{score}^{'t}_{g'} - \textit{score}^{t}_{g'}.
\end{equation*}
The improvement is evaluated as the difference of the grid-wise balance score measured pre-reallocation ($\textit{score}^{t}_{g'}$) and post-reallocation ($\textit{score}^{'t}_{g'}$). The reward is defined as,
\begin{align}\label{eq:reallocate_reward}
    r &= r_1 - r_2 + r_3; \\
    &= \widehat{SD_{g^c,t}} - \widehat{SD_{g'^c,t}} + \frac{1}{|G_N(g^c)|} \sum_{g' \in G_N(g^c)} \textit{score}^{'t}_{g'} - \textit{score}^{t}_{g'}.
\end{align}

\subsection{Architecture of the Strategic Dual-control Framework}
\label{subsect:dual-control}

The meal delivery system is a complex service system, consisting of multiple and various entities, including the couriers, orders, restaurants, and customers. In this research, our objective is to inspect how the interactions among these entities give rise to emergent patterns and dynamics at the system level. Meanwhile, we aim to investigate how local interaction rules can be learned to improve system-level performance. Given the inherent complexity of a meal delivery system, we have selected agent-based modeling \citep{macal2005tutorial} as the primary simulation tool to construct a digital twin environment for studying this system in our research. 

\begin{figure}[h]
    \centering
    \includegraphics[width= \textwidth]{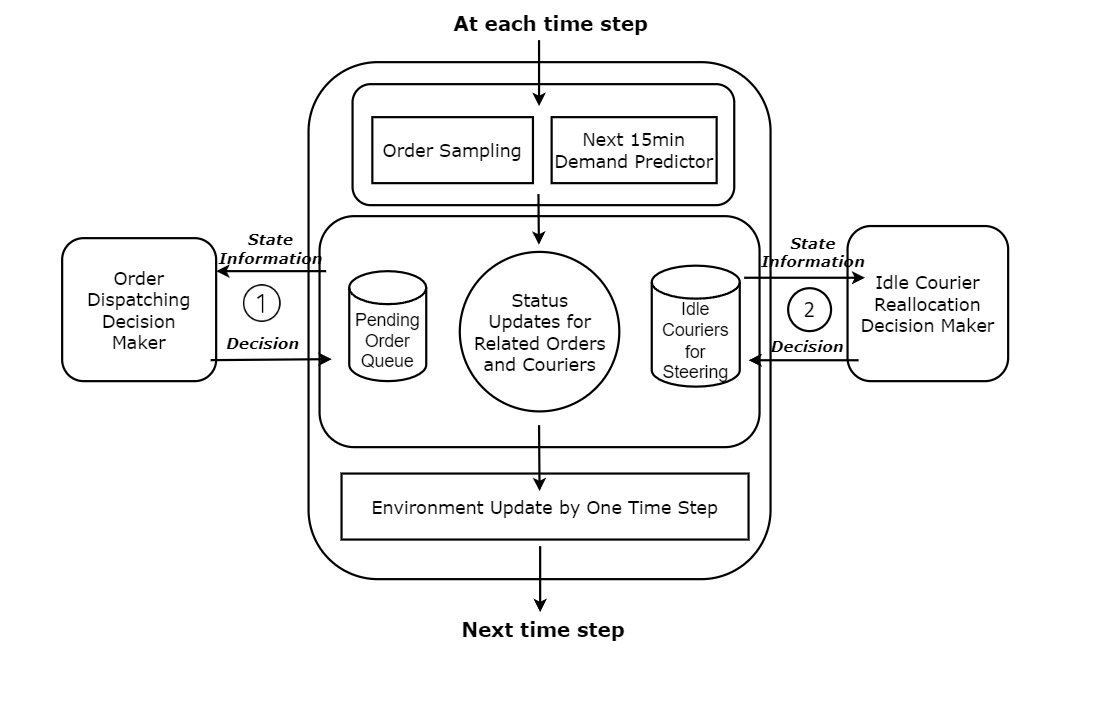}
    \caption{Simulation Process Diagram}
    \label{fig:diagram}
\end{figure}

Combining the definitions and assumptions presented earlier in this section, we propose an agent-based simulation framework for the meal delivery system, as depicted in Figure \ref{fig:diagram}. The framework contains three main components: the environment, the order dispatching decision generator, and the idle courier reallocation decision generator. The environment of the meal delivery system is responsible for maintaining the spatial distribution of pending orders and active couriers in the service network, updating the queue of pending order requests, and tracking the status of active orders and couriers. The environment is always firstly updated according to the last decision before starting the next decision-making process. Since the environment is always first updated according to the last decision before starting the next decision-making process, coordination between actions is achieved through the environment itself. 

At the start of each time step, new orders are sampled according to the specifications provided in Section \ref{subset:order}. The newly generated orders are then added to the queue for pending orders, awaiting assignment. 
Meanwhile, the forecasting model discussed in Section \ref{subsect:pred} is applied to predict the number of order arrivals for all restaurant-household grids in the upcoming 15 minutes. 
Proceeding to the stage of operation, the platform adopts the dual-control framework to sequentially generate real-time dispatching decisions for the pending order and then reallocate the idle couriers within the service network.
Firstly, the order dispatching phase is activated to sequentially dispatch the pending orders. 
Prior to decision-making, all the orders are ranked based on their urgency level in the order buffer. The urgency level of an order is higher if its remaining expected preparation time is shorter.
For each pending order in the queue, the platform generates the state vector by summarizing the related information in the environment. According to the state vector, a dispatching decision is made following the employed order dispatching policy, which is trained via the approach we describe in Section \ref{subsect:order_dispatching}. Then, the decision is executed by the platform. 
And the environment updates the status of the related order and courier involved in the executed dispatching action, and it moves on to the next order in the queue. 
When all the pending orders have been handled, the idle fleet steering phase begins, where reallocation instructions are provided to each of the idle couriers who have been waiting for a new task for more than five minutes.
Similar to the order dispatching phase, the platform generates a state vector based on the information of the target idle courier and the supply and demand information of the service network. The state vector is then sent to the idle fleet steering decision-maker. Once the reallocation decision is made following the trained idle courier reallocation policy and executed by the platform, the environment updates the information related to the target courier and then moves on to process the next awaiting courier in the queue. This process continues until reallocation instructions have been provided and executed for all the qualified idle couriers at the current time step. 
Finally, the timer is advanced by one time step for the environment, couriers, and orders. Each time step corresponds to one minute in practice. 
One simulation consists of 120 time steps corresponding to the two-hour shift from 19:00 to 21:00 on a Saturday evening.

%% file: Chapters/Ch5_Instances.tex
\subsection{Empirical Dataset}
\label{subsect:data}
Our experiments are conducted with the simulated order data provided by a third party for a pseudo city based on real-life order transaction data.
The original dataset comprises over 800,000 orders placed between 10:45 a.m. and 9:30 p.m from Monday to Sunday throughout 22 months. Appendix.\ref{Appendix_1} provides the exploratory data analysis we perform for the original data. Having analyzed the original dataset, we find the demand levels are higher around dinner time, particularly during the weekend. Furthermore, there are no significant trends, monthly or annual seasonality patterns found in the number of orders received per day. 

In this study, we consider a sample service region of size $5 \times 5$ grids in the city center. From the original data, we observe that grids with households are only located on the outskirts of the city, while the grids with restaurants are densely located in the city center. Hence, we only consider orders related to restaurants from the center nine-grid area of our selected sample service region. But all the grids within the sample service region contain households. The sample service area has the same layout as the visual illustration in Figure \ref{fig:example_process}. To further simplify the problem, we focus on studying the optimal decision policies for the two-hour shift from 19:00 to 21:00 on a regular Saturday evening.
Hence, we select the subset of transaction data that contains orders related to the sample service region during these two-hour Saturday evening shifts. 
A total number of 11128 orders are collected from 88 of such Saturday evening two-hour shifts in this data subset. Each entry of the transaction data contains the placement date and time, and the hashed grid addresses of the restaurant and household.  

\begin{figure}[h!]
     \centering
     \begin{subfigure}[b]{0.47\textwidth}
         \centering
         \includegraphics[width=\textwidth]{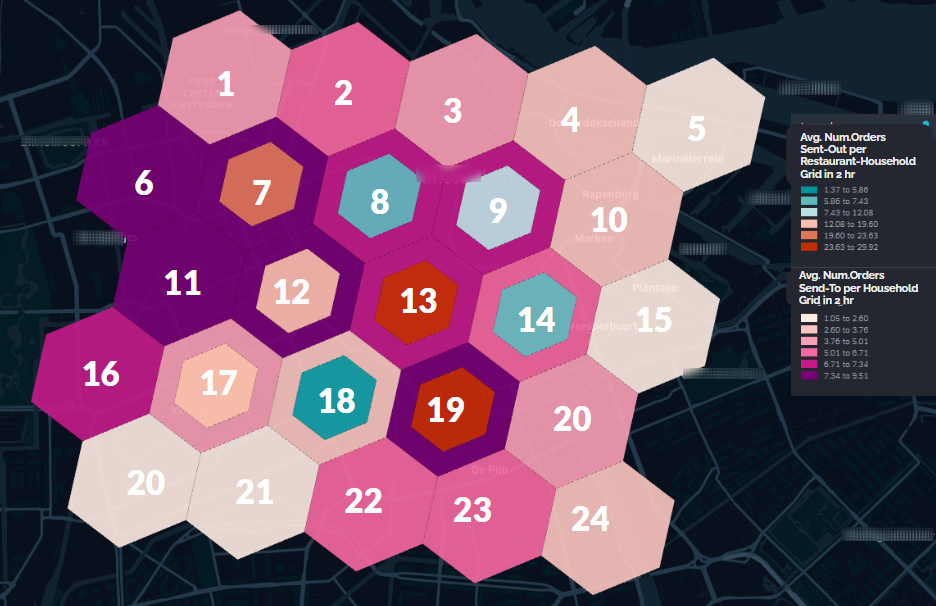}
         \caption{visualization of demand levels: (1) the average number of orders sent out from the restaurant-household grids (grids 7,8,9,12,13,14,17,18,19); (2) the average number of orders sent to the restaurant-household grids and household grids. The hotter colors represent higher demand.}
         \label{fig:map_rs_hh}
     \end{subfigure}
     \hfill
     \begin{subfigure}[b]{0.52\textwidth}
         \centering
         \includegraphics[width=\textwidth]{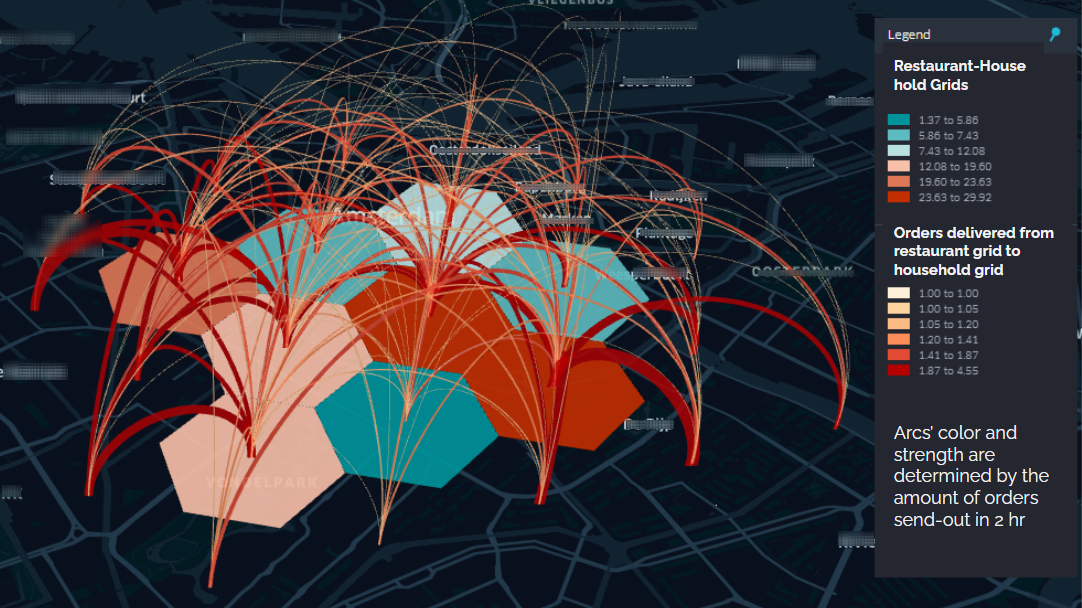}
         \caption{Visualization of the pairwise demand level between the origin grids (i.e., the grids with restaurants) and the destination grids (i.e., the grids with households). Each arc connects a pair of origin and destination grids. And the thickness and color of the arcs represent the pairwise demand level. The hotter color and thicker arc indicate higher demand.}
         \label{fig:map_rs_arc}
     \end{subfigure}
\caption{Visualization of the demand levels of the restaurant-household grids and household grids in our sample service region. The average number of orders is counted for those arrive at the sample service region during Saturday evening's two-hour shift (19:00 to 21:00) in the simulated data.}
\label{fig:h3map_visual}
\end{figure}

For grids that include restaurants, we can quantify their demand level by measuring the number of orders to be picked up/sent out from them as the origin of delivery within a two-hour shift. Similarly, for grids that consist of households, we can estimate the demand level by counting the number of orders sent to them as the destination. In Figure \ref{fig:map_rs_hh}, we visualize the sample service region using these two types of demand measures, where hotter colors indicate a comparatively higher demand level of the grids.
The central nine grids on the map with labels 7,8,9,12,13,14,17,18,19 are the restaurant-household grids, while the remaining grids are household-only grids. Among the nine grids with restaurants, grids 8, 14, and 18 exhibit relatively lower demand. To visualize the level of demand between origins and destinations, we represent the average number of orders sent out from a specific restaurant-household grid and delivered to a particular household grid using arcs in Figure \ref{fig:map_rs_arc}. 

\subsection{Policy Training}
For the training of order dispatching and idle courier steering policies, we propose a three-step \textit{sandwich learning strategy} for our RL-based strategic dual-control framework. And the training is performed via the simulated data generated from the agent-based simulator. 
In the first step, the order dispatching policy is learned by $R_1$ runs of simulation from an environment without idle courier steering. If the order dispatching policy has converged, we proceed to train the idle courier reallocation policy for $R_2$ runs with the previously trained order dispatching implemented in the system. If the policy hasn't converged yet, more runs should be performed till convergence is met. It is important to note that both RL-based policies cannot be trained simultaneously to avoid the problem of non-stationary environments. If more than one policy are changing in the environment, the moving target problem will make the learning challenging to stabilize and converge. Lastly, with the idle fleet steering policy implemented, we fine-tune the parameters of the order dispatching model by performing another $R_3$ runs of simulation. Within the random-access memory of 24GB, we recommend starting the training experiments with $R_1 = 200$, $R_2=150$, and $R_3=100$. 

\subsection{Performance Evaluation}

We evaluate the system performance of operational frameworks from three different aspects, covering the delivery performance efficiency in terms of time and courier usage efficiency, the analysis of supply and demand balancing within the service network, and the fairness of workload distribution among couriers. 

Time gap $\delta_o$ of a delivery is defined as the temporal difference between order ready and courier arrival times. It follows,
\begin{equation}\label{eq:time_gap}
    \delta_o = t^c_o - t^r_o,
\end{equation}
where $t^c_o$ is the arrival time of the courier and $t^r_o$ is the time when the order becomes ready. 
To assess the overall delivery efficiency, we evaluate the \textit{average time gap} between the order ready time and courier arrival time. 
A negative time gap indicates that there is no delay in delivery. Once the order is ready, the courier can collect it and start heading to the household. But it also implies a sunk cost of waiting time for the courier. On the other hand, a positive time gap means the order must wait for the courier to arrive before being sent out, which extends the delivery time and may lower customer satisfaction. Hence, we prefer the time gap to be negative while small in absolute value. The average time gap per run is taken as the average among all the orders delivered during the same shift. 

The \textit{average order pickup distance} assesses delivery efficiency by measuring the distance between the last active location of the assigned courier and the restaurant grid associated with the order. 
A shorter average pickup distance is desirable as it indicates lower operational costs for delivery in the long run and a more optimal distribution of couriers across the service network.

In this study, we assume an order is overdue if the system fails to assign the order within 10 minutes. Any overdue orders will be automatically canceled.
We measure the order fulfillment quality by the percentage of overdue orders among the orders generated, i.e., the \textit{overdue rate}.
The number of overdue orders captures the robustness of the system to provide high-quality delivery service. In the reward design of our order dispatching model, an overdue order is heavily penalized. The occurrence of overdue orders shows the extreme cases where the system becomes severely congested and runs short of couriers.

To keep track of the balancing of the spatial supply-demand distributions within the service network, we record the \textit{negative supply-demand balance scores}($NSD$), which is defined as the following,
\begin{align}
    NSD_t &= \sum_{g \in G} min\{N^C_{g,t} - N^O_g,0\}, \\
 \overline{NSD} &= \frac{1}{T} \sum_{t\in T} NSD_t
\end{align}
where $N^C_{g,t}$ denotes the number of couriers on grid $g$ at time $t$, $N^O_{g,t}$ the number of orders sampled from grid $g$ at time $t$. Note that the supply and demand gap is simply $N^C_{g,t} - N^O_g$.
The balance scores are averaged by the 120 time steps per shift. 
We prefer a close-to-zero value for average negative supply-demand balance score $\overline{NSD}$, as it indicates that the demand within the network can be promptly responded to by couriers within the same grid and under-supplied is less of an issue in this system. 

In order to assess the fairness in workload distribution among couriers, we conduct an analysis of several key statistics throughout a two-hour shift. The selected metrics include the \textit{delivery time} and \textit{idle time}, \textit{number of orders received}, and \textit{total distance traveled} of each courier. Note that the delivery time specifically accounts for the time spent on the order delivery tasks, excluding the time spent on reallocation tasks.
Then, we average these metrics among couriers of the same shift. Furthermore, we quantify the fairness of workload by reporting the standard deviations of these metrics across the couriers.  

To provide an objective evaluation of the frameworks, we split the provided simulated data into two parts, the evaluation and testing data. The evaluation data includes 8019 orders related to the sample service area we have selected. The historical transactions in evaluation data are collected from 63 randomly chosen two-hour shifts (19:00-21:00) on Saturday evenings. The remaining 25 two-hour shifts are adopted as the testing data in our experiments, containing 3151 orders. 
For the training of decision policies, we employed the simulated order data bootstrapped from the evaluation data, following the order sampling process described in section \ref{subset:order}.

To provide a comprehensive evaluation of the frameworks, we also report the performance over the simulated-evaluation data consisting of 100 randomly sampled two-hour shifts with our own simulator, next to the actual-evaluation data and the testing data respectively. By comparing the metrics obtained via the original order transactions, we can examine not only the performance difference of the trained policies comprehensively but also analyze the difference between the actual and simulated order streams.

%% file: Chapters/Ch6_Results.tex
In our experiments, we evaluate the performance of six different frameworks to assess and compare their performance in operating the meal delivery platform during a two-hour shift (19:00-21:00) on Saturday evening in the sample service region. The frameworks are categorized by the type of order dispatching policy adopted and whether idle courier reallocation policy is applied. It is important to clarify that for the myopic dual-control framework, the idle courier reallocation policy is also trained with current order and courier distributive information only. However, for all other cases, the strategic RL-based idle courier reallocation policies are generated using forward-looking information about future supply and demand distributions in the next 15 minutes.

Under the RL-based framework, we study the performance of strategic order dispatching policy trained with predicted demand and future courier distributive information, and the myopic dispatching policy trained with only current order and courier distributive information. Additionally, we also introduce a benchmark dispatching policy \textit{`Nearest Idle'}, which assigns orders to idle couriers closest to the restaurant grid, with tie-breaking done randomly. The benchmark policy follows first-come-first-served and postpones orders if no available idle courier is found. Similar `Nearest Idle' policies have been applied for real-time dispatching operations for ambulance \citep{jagtenberg2017dynamic} and on-demand ride-hailing \citep{feng2021we}. Our RL-based dispatching approach is more sophisticated than this benchmark policy from two perspectives. Firstly, our method considers the impact of assignment on supply-demand balance. Secondly, the dispatching decision made by our method aims to reduce the time difference between order ready time and courier arrival time. 

Before experimenting with the RL-based dual-control frameworks, we first examine the predictive accuracy of the lagged-dependent XGBoost (LD-XGBoost) method, which serves as the next 15-minute demand predictor in the framework providing demand anticipation for the service network. Our detailed evaluation is provided in Appendix.\ref{Appendix_3}, which suggests that the employed short-term demand forecasting method is capable of generating high-quality predictions.

In the remainder of this section, we present the computational results of different frameworks evaluated from four different aspects: computational time, delivery efficiency, supply-demand balance, and fairness of the workload distribution.
It is important to note that the reported averages and standard deviations are the mean values taken from the measurements per run. Outlier runs, determined by an average order time gap outside the $95^{th}$ percentile, are excluded from the report of performance.
To assess the significance of value differences between performance results, we apply the two-sided Mann-Whitney U test (Wilcoxon rank-sum test). The null hypothesis assumes identical distribution between the two groups of values. And we reject the null hypothesis if the p-value is less than a significance level of 0.05. Since the actual-evaluation and testing data sample sizes are relatively small, the statistical test is only conducted on performance results obtained from simulated-evaluation data.

\subsection{Computational Performance}
The average training and prediction generation times of different components in a dual-control framework are listed in Table \ref{tab:computation}. Accelerated by the gradient boosting feature of XGBoost, the training of LD-XGBoost predictor takes less than a half minute to accomplish for 6-month data. Also, results show that the demand prediction can be made in real-time without any problem.
Showed in Table \ref{tab:computation}, the computational time per learning update for both  RL-based dispatching and steering policies is rather short. Each batch learning takes less than a half second to accomplish with the acceleration of GPU. 
The average execution speeds of the trained order dispatching and idle courier reallocation policies are also very fast, with the generation of each decision taking less than 0.1 seconds. Hence, we believe the dual-control framework we propose can be implemented for real-time decision-making without any problem.
\input{Tables/Computational_Times}

\subsection{Overall delivery fulfillment efficiency and reliability}
Table \ref{tab:time_gaps} shows the average time gap per order. 
Except for the myopic dispatching without idle courier reallocation policy frameworks, the average time gaps of the other frameworks are consistently negative. It means, on average, the `click-to-door' delivery time of orders is not being delayed by late arrivals of couriers. 
Consistently, the utilization of an RL-based idle courier reallocation strategy has reduced the average time gap for the frameworks with all three kinds of order dispatching policies.
Moreover, for the frameworks with an RL-based order dispatching policy, the standard deviation of time gaps becomes lower when the idle courier reallocation policy is implemented. It indicates that the introduction of RL-based steering to the system is capable of not only improving but also stabilizing the time gap performance. The slightly higher variance for the `Nearest Idle' policy may be caused by fewer idle couriers available in the system due to  reallocation
\input{Tables/1-TimeGapTable}
Testing the statistical significance by the Mann-Whitney U tests, we find insignificant results only for the comparisons between the framework with strategic dispatching only and the framework with myopic dispatching and steering.

The average pickup distance per order results are listed in Table \ref{tab:pickup_distance}. Among the RL-based frameworks, the strategic dual-control framework with both order dispatching and idle courier reallocation policies has achieved the lowest average pickup distance and the smallest standard deviations with both evaluation and testing data. When only order dispatching is implemented, the framework with a strategic dispatching policy also has a significantly lower average pickup distance than the myopic case. 
Performing the Mann-Whitney U test among the results obtained by RL-based frameworks, we find that the average pickup distances are insignificant between the myopic frameworks with and without steering.

Since the frameworks with `Nearest Idle' dispatching always choose the available courier with the shortest distance to the restaurant grid, it is not reasonable to directly compare the average pickup distance across these benchmark frameworks. But we can still verify that the use of strategic idle courier reallocation has significantly reduced the pickup distance in both cases, meaning the courier distribution within the network is more efficient.  
\input{Tables/PickupDistance}

We have recorded the rates of overdue orders in Table \ref{tab:overdue}. During the simulated-evaluation experiments, overdue orders are found for the framework with myopic order dispatching only only. However, the rate of overdue increases significantly for the experiments with actual order transaction data, especially with the testing data, which was unseen by the models before. We believe this is caused by the difference in order arrival sequence between the simulated and actual data. Since the policies are also trained via the less realistic simulations, the general performance measured by other metrics for the RL-based frameworks is likely to be worse with the actual-evaluation and testing data.

Despite the flaws in simulation design, the rates of overdue orders show that the reliability of strategic frameworks is substantially higher than the myopic frameworks. Also, we notice clear decreases in overdue rates with implementing the idle courier reallocation policy for the framework with myopic dispatching.
\input{Tables/Overdue}

\subsection{Supply-demand balancing within the service network}
\label{subset:negative_score}
Recall that the average positive network score $\overline{PSD}$ measures the surplus, while the average negative network score $\overline{NSD}$ measures the deficit of idle couriers within the service network during the two-hour shift.
During the experiments, we find that the absolute values of the $\overline{PSD}$ are constantly larger than those of $\overline{NSD}$, indicating that the network is sufficiently supplied with 25 couriers. Hence, we mainly analyze the negative supply-demand balance scores to measure how often the supply deficit occurs. Table \ref{tab:negative_network_score} shows the $\overline{NSD}$ averaged per simulation. 
\input{Tables/NegativeNetworkScore}

Among the different frameworks, the strategic RL-based framework with both dispatching and steering policies generally obtains the least amount of deficit in supply on average. Nevertheless, its standard deviations are also the lowest compared to the other frameworks, meaning the low supply deficit performance is also the most stable. Furthermore, it is interesting to observe that the under-supplied situation measured by $\overline{NSD}$ becomes worse when a strategic reallocation policy is implemented for the `nearest idle' dispatching framework. It may be because fewer couriers are available for order assignments when some of them are on reallocation. 

According to the Mann-Whitney U tests, the distributions of average negative network supply-demand balance scores are not significantly different from each other among the values obtained from the frameworks that apply RL-based order dispatching policy.

\subsection{Fairness of workload distribution among couriers}
\label{subsect:fairness_results}
In Table \ref{tab:num_orders}, we compare the standard deviations of the number of orders received among couriers in the same shift. The standard deviations evaluated from frameworks with an RL-based dispatching policy are very close to each other but generally lower than those obtained from the frameworks with a `Nearest Idle' dispatching policy. It indicates that the fair embeddings of couriers and reactive rebalancing consideration of our RL-based dispatching policies fairly improve the chances of couriers getting selected. Furthermore, for the cases with `Nearest Idle' dispatching policy, the standard deviation values are reduced by implementing a strategic RL-based idle courier reallocation. It indicates that idle fleet steering helps increase the fairness of order assignment among couriers by reallocating the idle couriers back to the hub of demand at the center. However, the Mann-Whitney U tests suggest that the standard deviation values are not significantly different from each other. We should resort to other metrics to compare workload distribution fairness obtained by these frameworks. 
\input{Tables/OrdersReceived}

Table \ref{tab:on-task_time} shows the average and standard deviations of total delivery time (out of 120 minutes) per courier.
The average delivery time of couriers reflects the average usage of couriers during the two-hour shift. Hence, when the rate of overdue orders is zero, a lower average delivery time among couriers suggests that delivery planning is more efficient. Among the evaluation and testing experiments where a zero overdue rate is achieved, we find that the shortest average delivery time per courier is obtained by the framework that only implements the `nearest idle' dispatching policy. And the second lowest average delivery time per courier is achieved by the RL-based strategic dual-control framework with dispatching and steering. 
\input{Tables/Average_On-task_Time}

We inspect the standard deviations of delivery time in the shift to evaluate the task time fairness among couriers. Among the RL-based frameworks, the lowest standard deviation is obtained by the myopic dual-control framework with the simulated-evaluation data. In contrast, the strategic dual-control framework with the actual-evaluation and testing data achieves the lowest standard deviations. This evidence shows that implementing an RL-based idle courier reallocation policy enhances the fairness of delivery task workload distribution among couriers. 
The introduction of a strategic fleet steering policy has significantly reduced the variance of delivery time among couriers for the cases with the `Nearest Idle' dispatching policy, as shown by the Mann-Whitney U tests.

The last metric reported is the average total distance traveled per courier during a shift, listed in Table \ref{tab:total_distance}. Although it comes as a trade-off with an increase in the average total travel distance, the standard deviations have decreased significantly with the implementation of RL-based idle courier reallocation policies in the frameworks with all three different types of dispatching policies. For the experiment on simulated-evaluation data with the strategic frameworks, we noticed that incorporating courier reallocation policy only increases in average travel distance per courier is only 8\% (1.62 units of distance). But the reduction in standard deviation is up to 21\% compared to the case with strategic dispatching only, showing the idle courier reallocation greatly improves fairness.
The average total travel distances for `Nearest Idle' are shorter with moderate standard deviations among couriers. That is reasonable since the policy chooses the idle courier with the least distance to the restaurant grid. But the standard deviations have also been significantly reduced by 30.5\%, 38.1\% and 43\% respectively, after introducing the idle fleet steering policy.
For the average travel distance among couriers, the outcomes of Mann-Whitney U tests show that only the distributions between myopic dispatching without steering and `nearest idle' dispatching are not significantly different at a 95\% confidence level.
\input{Tables/TravelDistance}

%% file: Tables/Computational_Times.tex
\begin{table}[h!]
    \centering
    \caption{Computational Time for Each Components (unit in seconds).}
    \label{tab:computation}
        \begin{tabular}
        {T{0.32\columnwidth}T{0.32\columnwidth}T{0.30\columnwidth}}
        \hline
        \textsc{Model} &\textit{Training} & \textit{Execution}\\
        \hline
        \textbf{Next 15-min Demand Pred.} & 27.4 (\textit{sec} per training of a grid) & $2.8 \times 10^{-7}$ (\textit{sec} per prediction) \\
        \textbf{Order Dispatching} & 0.43 (\textit{sec} per learning update) & 0.065 (\textit{sec} per decision)\\
        \textbf{Idle Courier Reallocation} & 0.36 (\textit{sec} per learning update) & 0.057 (\textit{sec} per decision)\\
        \textbf{Nearest Idle} & - & $6.3 \times 10^{-5}$ (\textit{sec} per decision)\\
        \hline
        \multicolumn{3}{l}{Note: For the experiments related to the next 15 minute demand predictions, we utilize 6-month and } \\
        \multicolumn{3}{l}{1-month full order transaction data respectively for training and testing for all grids with restaurants}\\
        \multicolumn{3}{l}{in the city to show the predictor's capability in forecasting for the whole system throughout the day.}\\
        \multicolumn{3}{l}{}
        \end{tabular}
\end{table}

%% file: Tables/1-TimeGapTable.tex
\begin{table}[htbp]  
\caption{Time gaps between order ready time and courier arrival time.}
\label{tab:time_gaps}
  \begin{tabular}{T{0.20\textwidth}T{0.15\textwidth}T{0.06\textwidth}T{0.06\textwidth}T{0.06\textwidth}T{0.06\textwidth}T{0.06\textwidth}T{0.06\textwidth}}
    \hline \hline
    \textsc{Order} &  \textsc{Idle Courier} & \multicolumn{2}{c}{\textsc{Simu. Eva.}} & \multicolumn{2}{c}{\textsc{Actual Eva.}} & \multicolumn{2}{c}{\textsc{Testing}} \\
    \textsc{Dispatching}& \textsc{Reallocation}&  \textit{Avg.}& \textit{Std.} & \textit{Avg.}& \textit{Std.} & \textit{Avg.}& \textit{Std.}\\
    \hline
    \textbf{RL-based} & \textbf{-} & -1.25	& 5.04	& -1.93 &4.82&-1.35 &5.43\\
     \textbf{+ Strategic} & \textbf{RL-based}& -2.62 & \textbf{4.99}	&-3.19&	\textbf{4.52}&-2.67&	\textbf{5.16}\\
     \hline
    \textbf{RL-based } & \textbf{-} & 0.10&5.40&-1.02&5.03&-0.68&5.45\\
     \textbf{+ Myopic}& \textbf{RL-based} & -0.91& \textbf{5.09}& -1.97& \textbf{4.61}& -1.44& \textbf{4.98}\\ \hline
    \textbf{Nearest} & \textbf{-} &-4.86	&\textbf{3.38}&	-5.47&	\textbf{3.22}&	-5.49&	\textbf{3.24} \\
    \textbf{Idle} & \textbf{RL-based} & -5.02 & 3.39 & -5.72 & 3.29 & -5.66 & 3.38\\
    \hline \hline
    \multicolumn{8}{c}{\textsc{Notations}: \textbf{Simu.}: simulation, \textbf{Avg.}: average, \textbf{Std.}: standard deviation,} \\
    \multicolumn{8}{c}{\textbf{Eva.}: evaluation, \textbf{RL-based}: reinforcement learning-based}
  \end{tabular}
\end{table}

%% file: Tables/PickupDistance.tex
\begin{table}[htbp]  
\caption{Pickup distances of orders.}
\label{tab:pickup_distance}
  \begin{tabular}{T{0.20\textwidth}T{0.15\textwidth}T{0.06\textwidth}T{0.06\textwidth}T{0.06\textwidth}T{0.06\textwidth}T{0.06\textwidth}T{0.06\textwidth}}
    \hline \hline
    \textsc{Order} &  \textsc{Idle Courier} & \multicolumn{2}{c}{\textsc{Simu. Eva.}} & \multicolumn{2}{c}{\textsc{Actual Eva.}} & \multicolumn{2}{c}{\textsc{Testing}} \\
    \textsc{Dispatching}& \textsc{Reallocation}&  \textit{Avg.}& \textit{Std.} & \textit{Avg.}& \textit{Std.} & \textit{Avg.}& \textit{Std.}\\
    \hline
    \textbf{RL-based} & \textbf{-} & 1.78&	1.00&	1.60&	0.97&	1.63&	1.00 \\
     \textbf{+ Strategic}& \textbf{RL-based} & \textbf{1.38}&	\textbf{0.93}&	\textbf{1.21}&	\textbf{0.85}&	\textbf{1.20}&	\textbf{0.86} \\ \hline
    \textbf{RL-based} & \textbf{-} & 1.96&	1.02&	1.80&	1.00&	1.81&	1.00 \\
    \textbf{+ Myopic} & \textbf{RL-based} & 1.76&	0.97&	1.60&	0.94&	1.59&	0.94\\ \hline
    \textbf{Nearest} & - & 1.35	&0.85&	1.13&	0.78&	1.15&	0.80 \\
    \textbf{Idle}& \textbf{RL-based} & 1.28 &0.86 &1.05 & 0.81 & 1.04 & 0.79\\
    \hline \hline
    \multicolumn{8}{c}{\textsc{Notations}: \textbf{Simu.}: simulation, \textbf{Avg.}: average, \textbf{Std.}: standard deviation,} \\
    \multicolumn{8}{c}{\textbf{Eva.}: evaluation, \textbf{RL-based}: reinforcement learning-based}
  \end{tabular}
\end{table}

%% file: Tables/Overdue.tex
\begin{table}[htbp]  
\caption{Rates of Overdue Orders.}
\label{tab:overdue}
  \begin{tabular}{T{0.20\textwidth}T{0.15\textwidth}T{0.15\textwidth}T{0.15\textwidth}T{0.15\textwidth}}
    \hline \hline
    \textsc{Order} &  \textsc{Idle Courier} & \textsc{Simu. Eva.} & \textsc{Actual Eva.} &\textsc{Testing} \\
    \textsc{Dispatching}& \textsc{Reallocation}&  \textit{Rate(\%)} & \textit{Rate(\%)}& \textit{Rate(\%)}\\
    \hline
    \textbf{RL-based} & \textbf{-} & \textbf{0.00\%}	& 2.42\%	& 7.46\%\\
    \textbf{+ Strategic} & \textbf{RL-based} & \textbf{0.00\%}	&3.35\%	 &7.52\%  \\ \hline
    \textbf{RL-based} & \textbf{-} & 0.47\% & 14.17\% & 20.56\%  \\
    \textbf{+ Myopic} & \textbf{RL-based} & \textbf{0.00\%}	& 5.89\%	& 18.69\% \\ \hline
    \textbf{Nearest} & - & \textbf{0.00\%}&  \textbf{0.00\%}&  \textbf{0.00\%}\\
    \textbf{Idle}& \textbf{RL-based} & \textbf{0.00\%}& \textbf{0.00\%}& \textbf{0.00\%}\\
    \hline \hline
    \multicolumn{5}{c}{\textsc{Notations}: \textbf{Simu.}: simulation, \textbf{Eva.}: evaluation, \textbf{RL-based}: reinforcement learning-based.}
  \end{tabular}
\end{table}

%% file: Tables/NegativeNetworkScore.tex
\begin{table}[htbp]  
\caption{The negative supply-demand balance score within the network.}
\label{tab:negative_network_score}
  \begin{tabular}{T{0.15\textwidth}T{0.15\textwidth}T{0.07\textwidth}T{0.07\textwidth}T{0.07\textwidth}T{0.07\textwidth}T{0.07\textwidth}T{0.07\textwidth}}
    \hline \hline
    \textsc{Order} & \textsc{Idle Courier} & \multicolumn{2}{c}{\textsc{Simu. Eva.}} & \multicolumn{2}{c}{\textsc{Actual Eva.}} & \multicolumn{2}{c}{\textsc{Testing}} \\
    \textsc{Dispatching}& \textsc{Reallocation} &  \textit{Avg.}& \textit{Std.} & \textit{Avg.}& \textit{Std.} & \textit{Avg.}& \textit{Std.}\\
    \hline
    \textbf{RL-based} & \textbf{-} & \textbf{0.000}	&0.001	&-0.028	&0.035	&\textbf{-0.155}	&0.200\\
     \textbf{ + Strategic}& \textbf{RL-based} & \textbf{0.000}	&\textbf{0.000}	&\textbf{-0.010}	&\textbf{0.033}	&-0.177	& \textbf{0.197} \\ \hline
    \textbf{RL-based} & \textbf{-} & -0.002&	0.012&	-0.312&	0.404&	-0.401&	0.502 \\
    \textbf{+ Myopic} & \textbf{RL-based} & -0.020	&0.040	&-0.124	&0.146	&-0.329	&0.389 \\ \hline
    \textbf{Nearest} & \textbf{-} & -0.140	&0.440	&-0.140	&0.410	&-0.240	&0.550 \\
    \textbf{Idle}& \textbf{RL-based} & -0.173 & 0.558 &-0.163 & 0.484 & -0.259 & 0.659\\
    \hline \hline
    \multicolumn{8}{c}{\textsc{Notations}: \textbf{Simu.}: simulation, \textbf{Avg.}: average, \textbf{Std.}: standard deviation,} \\
    \multicolumn{8}{c}{\textbf{Eva.}: evaluation, \textbf{RL-based}: reinforcement learning-based}
  \end{tabular}
\end{table}

%% file: Tables/OrdersReceived.tex
\begin{table}[htbp]  
\caption{Standard deviations about the number of orders received by couriers from the same shift.}
\label{tab:num_orders}
  \begin{tabular}{T{0.15\textwidth}T{0.15\textwidth}T{0.15\textwidth}T{0.15\textwidth}T{0.15\textwidth}}
    \hline
    \textsc{Order} &  \textsc{Idle Courier} & \textsc{Simu. Eva.} & \textsc{Actual Eva.} & \textsc{Testing} \\
    \textsc{Dispatching}& \textsc{Reallocation}&  \textit{Std.} &  \textit{Std.} & \textit{Std.}\\
    \hline
    \textbf{RL-based} & \textbf{-} & 1.02	&1.16	&1.23  \\
    \textbf{+ Strategic}  & \textbf{RL-based} & 1.02	&\textbf{1.11}	&1.19  \\ \hline
    \textbf{RL-based} & \textbf{-} & \textbf{1.00}	&1.16	&\textbf{1.18} \\
    \textbf{+ Myopic} & \textbf{RL-based} & 1.03	&1.21	&1.19 \\ \hline
    \textbf{Nearest} & - & 1.25	&1.51	&1.55 \\
    \textbf{Idle}& \textbf{RL-based} & 1.04 & 1.19 & 1.23\\
    \hline
    \multicolumn{5}{c}{\textsc{Notations}: \textbf{Simu.}: simulation, \textbf{Std.}: standard deviation, \textbf{Eva.}: evaluation,} \\
    \multicolumn{5}{c}{\textbf{RL-based}: reinforcement learning-based.}
  \end{tabular}
\end{table}

%% file: Tables/Average_On-task_Time.tex
\begin{table}[htbp]  
\caption{The delivery time among 25 couriers.}
\label{tab:on-task_time}
  \begin{tabular}{T{0.15\textwidth}T{0.15\textwidth}T{0.07\textwidth}T{0.07\textwidth}T{0.07\textwidth}T{0.07\textwidth}T{0.07\textwidth}T{0.07\textwidth}}
    \hline \hline
    \textsc{Order} & \textsc{Idle Courier} & \multicolumn{2}{c}{\textsc{Simu. Eva.}} & \multicolumn{2}{c}{\textsc{Actual Eva.}} & \multicolumn{2}{c}{\textsc{Testing}} \\
    \textsc{Dispatching}& \textsc{Reallocation} &  \textit{Avg.}& \textit{Std.} & \textit{Avg.}& \textit{Std.} & \textit{Avg.}& \textit{Std.}\\
    \hline
    \textbf{RL-based} & \textbf{-} & 86.08	&14.64	&73.75	&16.21	&72.92	&16.94\\
     \textbf{+ Strategic}& \textbf{RL-based} & \textbf{83.71} &14.84	& \textbf{71.83}	& \textbf{14.85}	& \textbf{70.34} & \textbf{15.93} \\ \hline
    \textbf{RL-based} & \textbf{-} & 88.57	&14.37	&74.91	&15.88	&73.31	&16.29 \\
    \textbf{+ Myopic} & \textbf{RL-based} & 86.28	&\textbf{14.35}	&72.64	&16.05	&71.04	&16.23\\ \hline
    \textbf{Nearest} & - & \textbf{80.38}	&16.10	& \textbf{68.14}	&18.44	& \textbf{66.67}	&18.95 \\
    \textbf{Idle}& \textbf{RL-based} & 80.79& \textbf{13.58}& 68.57& \textbf{14.72} & 67.04 & \textbf{14.69} \\
    \hline \hline
    \multicolumn{8}{c}{\textsc{Notations}: \textbf{Simu.}: simulation, \textbf{Avg.}: average, \textbf{Std.}: standard deviation,} \\
    \multicolumn{8}{c}{\textbf{Eva.}: evaluation, \textbf{RL-based}: reinforcement learning-based}
  \end{tabular}
\end{table}

%% file: Tables/TravelDistance.tex
\begin{table}[htbp]  
\caption{The total travel distance per courier.}
\label{tab:total_distance}
  \begin{tabular}{T{0.20\textwidth}T{0.15\textwidth}T{0.06\textwidth}T{0.06\textwidth}T{0.06\textwidth}T{0.06\textwidth}T{0.06\textwidth}T{0.06\textwidth}}
    \hline
    \textsc{Order} &  \textsc{Idle Courier} & \multicolumn{2}{c}{\textsc{Simu. Eva.}} & \multicolumn{2}{c}{\textsc{Actual Eva.}} & \multicolumn{2}{c}{\textsc{Testing}} \\
    \textsc{Dispatching}& \textsc{Reallocation}&  \textit{Avg.}& \textit{Std.} & \textit{Avg.}& \textit{Std.} & \textit{Avg.}& \textit{Std.}\\
    \hline
    \textbf{RL-based} & \textbf{-} & 20.49	&4.21	&15.16	&4.06	&14.99	&4.41 \\
    \textbf{+ Strategic} & \textbf{RL-based} & 22.11	&\textbf{3.31}	&18.42	&\textbf{2.55}	&18.36	&\textbf{2.52}\\ \hline
    \textbf{RL-based} & \textbf{-} & 21.51&	4.44&	16.05&	4.11&	15.82&	4.24 \\
     \textbf{ + Myopic} & \textbf{RL-based} & 24.09	&\textbf{3.14}	&20.26	&\textbf{2.57}	&20.15	&\textbf{2.59} \\ \hline
    \textbf{Nearest} & - & 17.51	&4.34	&12.23	&4.44	&12.31	&4.15 \\
    \textbf{Courier} & \textbf{RL-based} & 19.02	&\textbf{2.73}	&15.21	&\textbf{2.22}	&15.27	&\textbf{2.15} \\ \hline
    \textbf{Nearest} & - & 18.04	&3.77	&12.68	&3.57	&12.37	&3.65 \\
    \textbf{Idle}& \textbf{RL-based} & 21.39 & \textbf{2.62} & 17.35 &\textbf{2.21} & 17.28 & \textbf{2.08}\\
    \hline
    \multicolumn{8}{c}{\textsc{Notations}: \textbf{Simu.}: simulation, \textbf{Avg.}: average, \textbf{std.}: standard deviation,} \\
    \multicolumn{8}{c}{\textbf{Eva.}: evaluation, \textbf{RL-based}: reinforcement learning-based}
  \end{tabular}
\end{table}

%% file: Chapters/Ch7_Discussion.tex
Among the RL-based operation framework, we have observed that, in most experiments, the delivery performance of the strategic frameworks with forward-looking information has outperformed the myopic frameworks, which do not benefit from the anticipatory demand information. This evidence of superior performance in delivery efficiency and reliability, supply-demand balancing, and workload distribution fairness suggests that the incorporation of high-quality short-term demand predictions can significantly improve the quality of decision-making in the system. The success of our RL-based strategic dual-control framework underscores the potential of the predict-then-optimize approach, emphasizing how explicit predictions can further enhance the performance of strategic decision-making with a foresighted optimization algorithm like reinforcement learning.

Focusing on the comparison of the RL-based order dispatching policies to the `Nearest Idle' benchmark, we notice that the average waiting times of couriers at the restaurant are lower via the RL-based dispatching framework. The reduction in wasted waiting time at the restaurant reflects the consideration of enhancing the courier's time efficiency as part of the multi-objective optimization problem. 
Additionally, results discussed in Section \ref{subset:negative_score}  indicate that the under-supplied condition has  been significantly improved compared to the cases with benchmark dispatching policy. Among the performance of dispatching-only frameworks, we find that the standard deviations are significantly lower for the number of orders received and delivery time per courier in Section \ref{subsect:fairness_results}.
These observations suggest that our order dispatching policy's passive rebalancing consideration contributes to better matching of network supply-demand distributions by selecting couriers from oversupplied areas. 

Our experiments have shown that idle courier steering policies are helpful in improving delivery efficiency for the frameworks with RL-based dispatching policies. At the evaluation of average pickup distance for orders, the results obtained by the strategic frameworks have become significantly lower with the use of an idle courier reallocation policy, especially for the cases with strategic controls. And the average time spent on delivery is  shorter per courier for a similar amount of work. This evidence suggests the distribution of couriers within the service network has become more adaptive to the dynamics of demand with the application of the self-organized and proactive supply steering in the system.
Nevertheless, the strategic idle fleet steering has  significantly enhanced the fairness of workload distribution among couriers. By analyzing 
the standard deviations in the delivery time, number of orders received, and the total distance traveled per courier in Section \ref{subsect:fairness_results}, we believe the implementation of the reallocation policy reduces assignment biases among couriers by moving couriers at less-advantaged locations back to the area with more delivery opportunities. Considering both benefits of idle courier reallocation, we believe it is a beneficial solution to incorporate the RL-based self-organized supply steering approach as part of the automated operation process of the on-demand meal delivery system.

Jointly considering the advantages of strategic decision-making with predicted demand, RL-based order dispatching and idle courier reallocation components, we believe the strategic RL-based dual-control framework offers a promising approach for self-organized operation in on-demand meal delivery. While this framework may not outperform in every evaluated aspect, it demonstrates the potential of its multi-objective design to effectively balance different objectives.

%% file: Chapters/Ch8_Conclusion.tex
In this study, we intend to show that high-quality delivery performance of a meal delivery platform can be achieved by operating hourly-paid fleets and providing full instructions to the couriers. With these operation assumptions, we model the decision-making from the platform perspective, which lets the platform carry the risk of operational failure while reserving operational control over the system for the operator. 
To find a solution to these operational challenges, we investigate the possibility of using a reinforcement learning-based framework to find order dispatching and idle courier reallocation policies which automate the real-time operation process while simultaneously optimizing the delivery performance, courier usage efficiency, and improving fairness among couriers at the network-level.
Using a synthesis dataset, we construct the strategic dual-control framework with agent-based modeling techniques, which is used as the digital twin of the meal delivery system for policy training and model evaluation. In order to generate forward-looking decisions, we follow a `predict-then-optimize' paradigm by applying a short-term demand forecasting algorithm to provide predictive information for the RL-based strategic policies. Using Double Deep Q Networks (DDQN) as the basic RL algorithm, we design the strategic dual-control framework with order dispatching and idle courier reallocation. For experiments, we assume a fleet of 25 couriers is scheduled for a two-hour shift from 19:00 to 21:00 on a Saturday evening to serve the sample service area with 9 restaurant-household grids and 25 household-only grids. 

The experiment results show that our proposed framework is able to generate decisions in real-time with our RL-based strategic dual control framework.
Also, most orders can be picked up right after they are served from the kitchen. And the average courier waiting time at the restaurant is reduced compared to the case with industrial benchmark dispatching. Additionally, the under-supplied situation within the service network is alleviated using our approach, compared to the cases with benchmark order dispatching policy.
According to our analysis of the average on-task time, total travel distance, and the number of orders received, the workload distribution has also become more balanced with the use of our strategic dual-control framework. These findings indicate that the system-level delivery service quality remains high while the assignment fairness among couriers is improved with the application of our proposed framework. 
Methodologically, we show that a strategic policy including forward-looking demand predictive information improves the performance compared to a policy with only myopic information generated from the currently known demand in the system. Moreover, via the success of the decentralized decision-making of idle courier reallocation policy with local neighborhood information, we show that operations can be done at the local level but still achieve system control at the network level.
We believe these achievements of our RL-based strategic dual-control framework can inspire the operation system design of other on-demand services in the industry.

Our study can be further improved on the following aspects.
In this study, we have to consider not only algorithm and learning-related hyperparameters, such as the learning rate and experience replay buffer size, but also the reward-related hyperparameters, including the weight parameters for the multi-objective reward function for order dispatching. With CPU (GTX 1650) acceleration enabled, each training for the strategic dual-control framework takes about 3 hours, while the performance evaluation using simulated-evaluation data takes around 45 minutes to complete.
Limited by the time budget, we were unable to perform a systematic and thorough fine-tuning of the hyperparameters for our proposed framework. The values of hyperparameters are selected after conducting several trials and attempts. In order to achieve better performance, it is necessary to conduct further experiments to fine-tune the hyperparameters.
Due to computation constraints from simulation, another limitation of this study is that the selected service network in our experiments is rather small compared to the practice. In real life, a city may cover more than 100 grids. In the future, the simulation may be improved by parallel computing techniques in order to extend the experiment to a large-scale network. 

The rapidly growing on-demand services have recently become an active yet challenging field of research for operations research and urban transportation science. These services present dynamic and stochastic system properties and their need for real-time and forward-looking decision-making. It will be interesting to extend our framework to other on-demand applications with different system characteristics and operation challenges.
Having shown the benefit of the use of demand predictions in training strategic RL-based policies, we also suggest to extend our framework with other prediction methods for future research. For instance, instead of grid-wise demand forecasting, the spatial demand distribution can be predicted via Graph-based Long short-term memory networks (LSTMs) \citep{davis2020grids}. Continuous learning with RL has shown great adaptability to dynamic and fast-changing environments \citep{gu2016continuous,nagabandi2018deep}. Future research can also extend our framework to allow online learning of decision policies, and examine whether it is applicable and desirable to employ adaptive dispatching and steering policies in practice.

%% file: Chapters/Appendix_1.tex
\begin{figure}[h]
    \centering
    \includegraphics[width=0.5\textwidth]{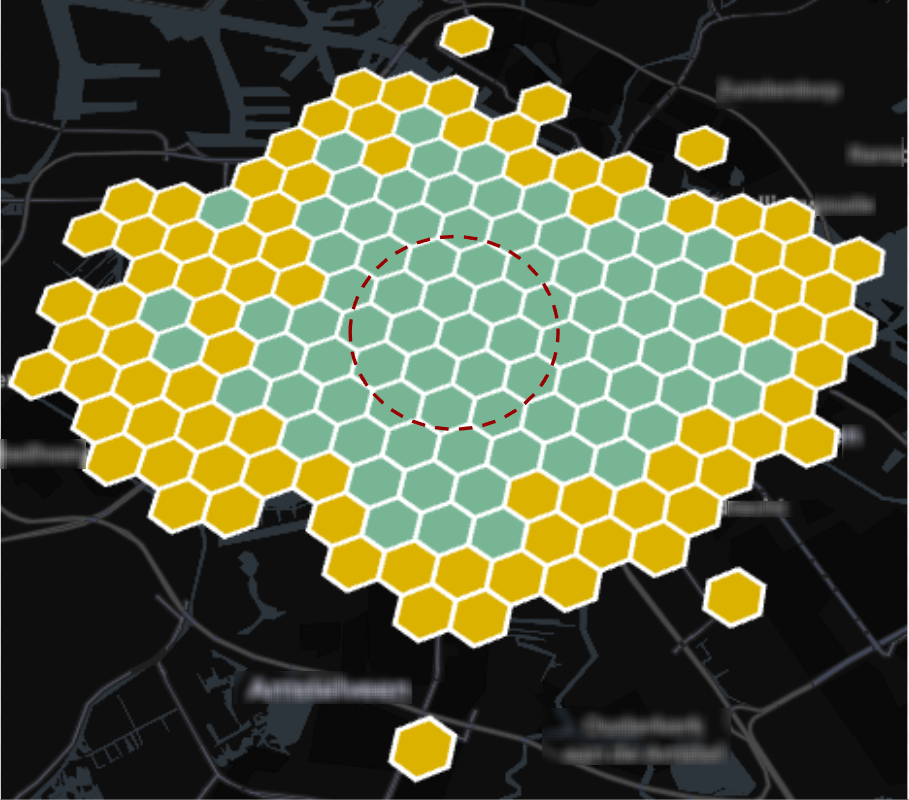}
    \caption{Distribution of restaurant-household grids and household grids in the example city. There are in total 50 restaurant-household grids on the map, which are colored in green. There are 92 household grids, colored in yellow. In this study, we investigate a smaller service region sampled from the city, which is denoted by the dashed circle at the center of the service map.}
    \label{fig:ams_full_map}
\end{figure}

In this section, we present the exploratory data analysis for the original simulated order transaction data. The geographical distribution of the restaurant-household grids and household grids within the service region of the city is visualized in Figure \ref{fig:ams_full_map}. It is important to note that orders can be picked up from and delivered to restaurant-household grids, as these grids contain both restaurants and households. On the other hand, household grids consist of households only, making them possible destinations for order deliveries, but not pick-up locations. The example city's service area consists of 50 restaurant-household grids and 92 household grids. As visualized in Figure \ref{fig:ams_full_map}, the restaurant-household grids are centrally located in the city, surrounded by household grids on the outskirt. While there may be a few household grids that are not directly adjacent to the others, the overall service area is well-connected. 

\begin{figure}[h!]
    \centering
    \includegraphics[width=0.70\textwidth]{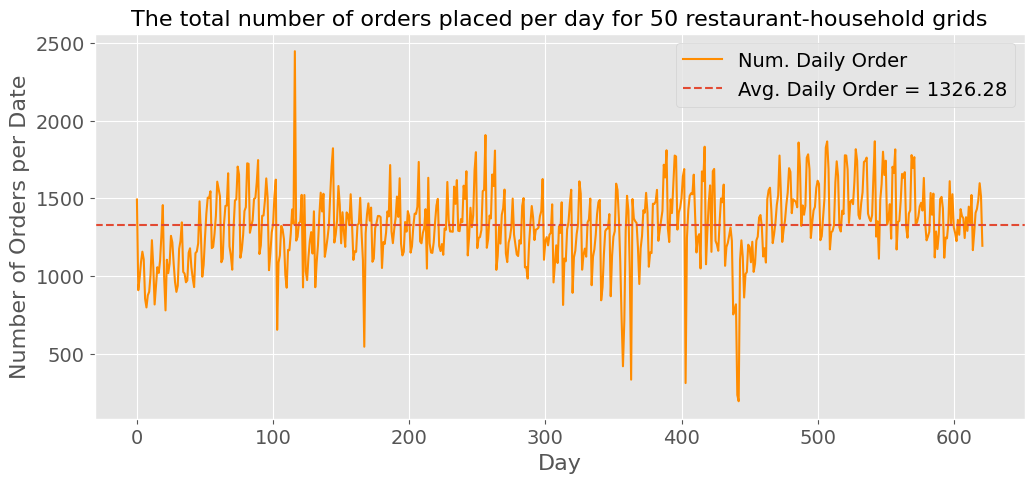}
    \caption{The number of orders received per date in the city, from month 1 to month 22.}
    \label{fig:ams_daily_order}
\end{figure}

\begin{figure}[h!]
    \centering
    \includegraphics[width= 0.70 \textwidth]{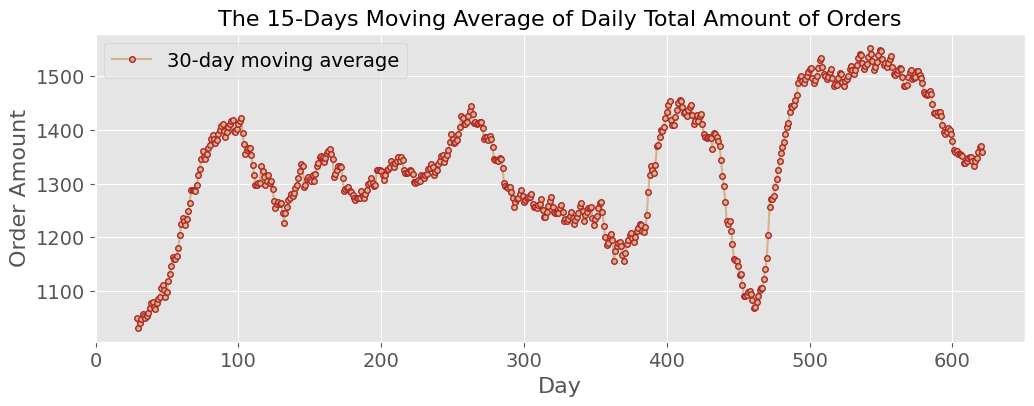}
    \caption{The simple 15-day moving average series of the number of orders received per date in the city, from month 1 to month 22.}
    \label{fig:rolling_window}
\end{figure}

To explore the seasonal demand patterns, we aggregate the orders placed on the same date. The numbers of orders placed per date are plotted in Figure \ref{fig:ams_daily_order}, revealing a weekly seasonal pattern. In this daily order sequence, we constantly observe small spikes followed by small dips in demand. Overall, the demand level remained rather stable during the period of 22 months, except for some unexpected peaks and plunges. To better understand the potential trends and seasonal (e.g., monthly) patterns in demand, we analyze daily order data by computing its simple moving averages, which helps to smooth out fluctuations. 
The simple moving average of time window $i$ is given by the following equation, $SMA_i = \frac{1}{N} \sum^{i}_{d=i-N+1} x_d$, where $x_d$ is the number of orders received on date $d$.
Applying a rolling window of 15 days, we compute the simple moving averages of the number of orders placed per day, which is plotted in Figure \ref{fig:rolling_window}. We find no obvious monthly or annual patterns in the visualized sequence of moving averages.

\begin{figure}[h!]
    \centering
    \includegraphics[width= 0.70 \textwidth]{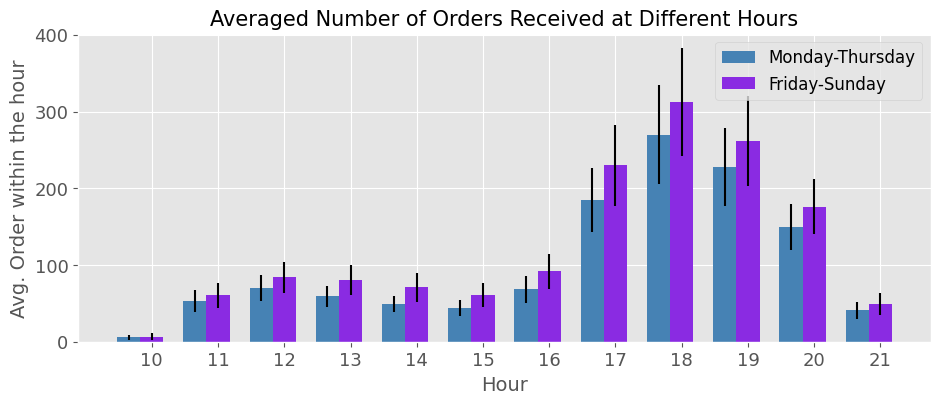}
    \caption{The average number of orders placed during different hours of the day, differentiating between Monday to Thursday and Friday to Sunday.}
    \label{fig:hourly_orders}
\end{figure}

Friday to Sunday are typically the busiest days for businesses in a week \citep{kimes2004restaurant}. Hence, the transaction data can be split into two groups: Monday to Thursday, and Friday to Sunday. To explore the variations of demand throughout the day, we compare the average number of orders received per hour on a day of these two groups in Figure \ref{fig:hourly_orders}. The hourly demand patterns for both groups reveal an asymmetric dual-peak pattern. The first peak is relatively moderate, which occurs around lunchtime around 12:00. The second peak of orders is comparatively high, which is observed during dinnertime between 17:00 and 21:00. Additionally, the average demand per hour is generally higher on a day from Friday to Sunday compared to a day from Monday to Thursday.

The analysis of demand patterns suggests that the hour of the day, as well as the day of the week are potentially good temporal features for demand predictions. Furthermore, our analysis indicates that the busiest period of the week occurs during the dinner hours from Friday to Sunday.

%% file: Chapters/Appendix_2.tex
We review the essentials and basics of reinforcement learning in this section, focusing on the related knowledge for the design of our RL-based operation framework. For further details of RL algorithms, we recommend the literature by \citep{sutton2018reinforcement}.

Reinforcement learning is a learning algorithm that is able to solve the decision problems that can be described in a Markov Decision Process (MDP) framework by directly interacting with the system. 
In an MDP, a decision agent obtains a \textit{state} $s$ (when full information is available about the environment) or an observation (when the environment is partially observable for the agent) from the \textit{environment} and selects an \textit{action} $a$ according to its decision \textit{policy} $\pi(s)$, or $\pi(a|s)$ stochastic decisions are assumed ($\sum_a \pi(a|s)=1$). Note that the state space $\mathcal{S}$ and action space $\mathcal{A}$ can be either discrete or continuous, depending on the problem.
The term `environment' in RL refers to the system where the agent interacts by executing actions and receiving immediate \textit{rewards} $r$ equals to $R(s,a)$. To accelerate training and reduce computational costs, simulated systems are often employed instead of directly working with real-world systems. 
Upon the execution of an action, the environment state evolves to a new state $s'$. The state transition probabilities are given by $\mathcal{P}(s'|s,a)$. 
Each interaction in the system can therefore be described by $(s,a,r,s')$.
And an episode in reinforcement learning refers to the full sequence of interactions from the starting state $s_0$ to termination. In some cases, a terminal state $s_T$ is defined. Reaching the terminal state indicates the termination of an episode. For instance, drawing a straight line in Tic Tac Toe will automatically win the game, and the episode terminates when the straight line appears.
However, many empirical applications are often continuous, such as meal delivery dispatching and autonomous driving. For continuous/non-episodic problems, there is no natural terminal state, and in the simulation, we often just terminate the run with it reaches the maximum number of time steps $T$. 
The goal of the agent is to maximize the cumulative discounted return $G$, which normally includes a discount factor $\gamma \in [0,1)$ to represent the diminishing values of future rewards. The cumulative discounted return $G_t$ at time step $t$ is given by, 
\begin{equation} \label{eq:cumulative_return}
    G_t = \sum^T_{k=0} \gamma^k R_{t+k+1}.
\end{equation}

State value functions $V_{\pi}$ are introduce to describe the estimated quality of state $s$ following the policy $\pi$, which is formulated by 
\begin{equation} \label{eq:state_value}
    V_{\pi}(s) \doteq \mathbf{E}_{\pi} [G_t|S_t=s] = \mathbf{E}_{\pi} [\sum^T_{k=0} \gamma^k R_{t+k+1}|S_t=s].
\end{equation}
We can obtain the following equality relation,
\begin{equation}
    V_{\pi}(s) = \mathbf{E}_{\pi} [R_{t} + \gamma V(s')]
\end{equation}
Connecting to the state transition probabilities, the state value function can be rewritten as,
\begin{equation*} 
    V_{\pi}(s) = \mathbf{E}_{\pi} [G_t|S_t=s] = \sum_{a \in A} \pi(a|s) \sum_{s',r} \mathcal{P}(s',r|s,a) (r + \gamma V_{\pi}(s'))
\end{equation*}    
For a given state-action pair, the \textbf{state-action value function} $Q$ estimates the expected return by,
\begin{equation} \label{eq:q_func}
    Q_{\pi}(s,a) \doteq \mathbf{E}_{\pi} [G_t|S_t=s,A_t=a] = \mathbf{E}_{\pi} [\sum^T_{k=0} \gamma^k R_{t+k+1}|S_t=s, A_t=a]
\end{equation}
By Bellman optimally equations, when values $V$ and $Q$ are optimal, it satisfies,
\begin{align*}
    V^*(s) &= max_a Q^*(s,a), \\
    Q^*(s,a) &=  \mathbf{E} [R_{t+1} + \gamma max_{a'} (S_{t+1},a')|S_t=s,A_t=a].
\end{align*}
Since the rewards and state transition probabilities are unknown at the beginning, value-based RL algorithms aim to iteratively update their estimation of value functions by interactions while searching for the optimal policy based on the value estimates \citep{sutton2018reinforcement}.
Tabular method Q-learning (see Algorithm \ref{alg:QL}) is introduced to learn the optimal $Q^*(s,a)$ via temporal-difference(TD) learning, which means the updates are performed on the previous samples and estimations. Q-learning is an off-learning approach, which means the updated target policy (greedy) is different from the behavior policy for evaluation (\textepsilon-greedy) where exploration exists. 
\input{Algorithms/Q-learning}

However, maximization bias is a problem in vanilla Q-learning caused by the bootstrapped updates over the same set of state-action value estimates. It leads to the problem that the overestimated values are more likely to be sampled, while the underestimated ones are harder to be corrected. To solve this issue, double Q-learning \citep{hasselt2010double} is introduced to mitigate the selection bias by maintaining two sets of Q value estimates, where the action selection and estimation update are performed with separate groups of estimates at each time.

The development of deep learning (DL) enables non-linear approximations of Q values by employing neural networks. Benefiting from the recent advancement in DL, Deep Reinforcement Learning (DRL) gives access to the incorporation of complex state inputs such as images and text data. Deep Q-Network(DQN) is introduced by DeepMind in 2015 \citep{mnih2015human}. However, the `deadly triad issue' becomes a problem when bootstrapping, off-policy, and function approximations are used together, which can potentially cause divergence and instability during training \citep{sutton2018reinforcement}.
There are two common techniques to enhance the learning stability of DQN. \textit{Experience replay} employs a memory buffer to store the previous experience $(s,a,r,s')$. By learning from a batch of randomly sampled experiences from the memory buffer in DQN, the model does not overfit to recent experience, and the variance of value function updates is decreased. As the learning proceeds or the environment evolves, the memory buffer is able to be updated adaptively by replacing the old experiences with new ones. The size of the memory buffer determined the maximum number of transitions stored. Similar to the idea of double Q-learning, the other technique is the application of a separate \textit{target network} next to a value network to smooth out the learning process. To prevent the moving target problem, the Q values of the future states are evaluated based on the target network, while the learning updates and action selection are performed with the value network. Every fixed period of time, the target network is updated by copying the parameter values from the value network. A DQN that implements a target network structure is referred to as \textit{Double Deep Q Network} (DDQN), which we utilize as the basic RL algorithm for our framework. 

There are other techniques to improve training stability and accelerate learning. Value clipping techniques, including reward clipping, gradient clipping, and error clipping, are sometimes employed to reduce the update step size. In this study, we clip the gradients to values between -0.5 and 0.5 to stabilize learning updates. The other trick we apply is epsilon decay for controlling the exploration-exploitation in action selection. 
Exploration at the beginning of the learning process is important for the agent to build up knowledge about the environment. On the flip side, exploitation is more important to maximize the reward as the value estimations approach convergence. In our DDQN algorithm, we employ the following epsilon-decay scheme,
\begin{equation} \label{eq:epsilon_scheme}
    \varepsilon = max\{ \kappa ^{T_L} \cdot \varepsilon_{0}, \varepsilon_{min}\},
\end{equation}
where $T_L$ denotes the total times that the algorithm has learned so far, $\kappa$ is the decay vector, $\varepsilon_0$ is the starting exploration vector value, $\varepsilon_{min}$ is the minimize exploration vector value.

\input{Algorithms/DDQN}

Algorithm \ref{alg:DDQN} illustrates the design of our DDQN method. For the meal delivery operating problems during a two-hour shift, the terminal states $s_T$ are defined to be the states obtained when $T=120$.
To clarify the terminology, we now describe an episode as a run of simulation.

The hyperparameters we select for the DDQN training for both order dispatching and idle courier reallocation policies are presented in Table \ref{tab:hyper_param}.

\input{Tables/HyperParameters}
During experiments, we find that the model is relatively sensitive to update. Therefore, we adopt a rather small learning rate.
And the Adam optimizer is applied with the clipped gradient values. The replay buffer size is set to a relatively limited value. That is because the dynamics of the simulation change a lot as the policy is learned and updated. Hence, the replay buffer should also be more adaptive to new changes to improve the efficiency of learning.

%% file: Algorithms/Q-learning.tex
\begin{algorithm}
\caption{Q-learning (off-policy TD control) for estimating $\pi \approx \pi_*$}
\label{alg:QL}
\begin{algorithmic}[1]
\Require Learning rate $\alpha \in (0, 1]$, exploration factor $\epsilon \in (0, 1)$ for $\varepsilon$-greedy action selection
\Ensure $Q(s, a)$ is initialized for all $s \in \mathcal{S}$, $a \in \mathcal{A}(s)$, except $Q(s_T, \cdot) = 0$ for terminal state $s_T$

\For{episode = 1 to $M$}
    \State Initialize environment state $s$
    \For{$t = 1$ to $T$}
        \State Choose action $a$ from state $s$ using policy derived from $Q$ (e.g., $\varepsilon$-greedy)
        \State Take action $a$, observe reward $r$, and next state $s'$
        
        \State Update Q-value for the current state-action pair:
        \State \quad $Q(s, a) \Leftarrow Q(s, a) + \alpha \left[ r + \gamma \max_{a' \in A(s')} Q(s', a') - Q(s, a) \right]$
        
        \State Update state: $s \Leftarrow s'$
    \EndFor
\EndFor
\end{algorithmic}
\end{algorithm}

%% file: Algorithms/DDQN.tex
\begin{algorithm}[H]
\caption{DDQN Algorithm with Experience Replay and Epsilon Decay, updated by Adam optimizer}
\label{alg:DDQN}
\begin{algorithmic}[1]
\Require \\
Initialize replay memory buffer $\mathcal{MB}$ with capacity $N$ \\
Initialize target network $Q^-$ and value network $Q$ with parameters $\theta^- \leftarrow \theta_0$, $\theta \leftarrow \theta_0$ \\
Set exploration rate $\varepsilon \leftarrow \varepsilon_0$
\Ensure Update the parameters using Adam optimizer

\For{episode = 1 to $M$}
    \For{t = 1 to $T$}
        \State Choose action $a$ using $\varepsilon$-greedy policy based on $Q(s,\cdot)$
        \State Observe reward $r$ and next state $s'$
        \If{$t = T$}
            \State $\textit{done} \Leftarrow 1$
        \Else
            \State $\textit{done} \Leftarrow 0$
        \EndIf
        \State Store transition $(s, a, r, s',\textit{done})$ in memory buffer $\mathcal{MB}$
        \State Update state $s \Leftarrow s'$
        
        \If{Learning condition met}
            \State Randomly sample minibatch of transitions from $\mathcal{MB}$
            \For{each transition $(s_i, a_i, r_i, s'_i, \textit{done}_i)$}
                \If{$\textit{done}_i \neq 1$}
                    \State Calculate target $y_i \Leftarrow r_i + \gamma \max_{a'} Q^-(s'_i, a'; \theta^-)$
                \Else
                    \State $y_i \Leftarrow r_i$
                \EndIf
            \EndFor
            \State Update Q-network by minimizing loss:
            \State $\mathcal{L}(\theta) = \frac{1}{N} \sum_{i} \left( y_i - Q(s_i, a_i; \theta) \right)^2$
            \State Perform stochastic gradient descent on $\mathcal{L}(\theta)$ using Adam
            \State Update value network parameters $\theta \Leftarrow \alpha \theta + (1 - \tau) \theta$
            \State Decay exploration rate $\varepsilon \Leftarrow \varepsilon \cdot \kappa$
        \EndIf
        
        \If{every $C$ steps}
            \State Update target network: $\theta^- \Leftarrow \theta$
        \EndIf
    \EndFor
\EndFor
\end{algorithmic}
\end{algorithm}

%% file: Tables/HyperParameters.tex
\begin{table}[h!]
    \centering
    \caption{The chosen hyperparameter values for the training of Conv-DDQN and DDQN algorithms.}
    \label{tab:hyper_param}
\begin{tabular}
{|F{0.2\textwidth}F{0.2\textwidth}|F{0.25\textwidth}F{0.20\textwidth}|}
    \hline
    \textsc{Learning Related Hyperparameters} & \textsc{Specification} & \textsc{Algorithm Related Hyperparameters} & \textsc{Specification} \\
    \hline 
    \textbf{Learning Rate} & 0.0005 & \textbf{Size of Replay Buffer} & 1000\\
    \hline
    \textbf{Discount Factor} & 0.8 &\textbf{Target Network Update Frequency} & per 100 decisions\\
    \hline
    \textbf{Exploration Scheme} &  $\kappa=0.99, \varepsilon_0 = 0.95, \varepsilon_{min} = 0.005$ for Eq \eqref{eq:epsilon_scheme}&\textbf{Batch Size} & 300\\
    \hline
    \textbf{Optimizer} & Adam (a stochastic gradient descent method)\citep{kingma2014adam} & \textbf{Learning Frequency} & Every 5 time steps\\
    \hline
\end{tabular}
\end{table}

%% file: Chapters/Appendix_3.tex
To assess the short-term demand forecasting model's predicting capability in a more comprehensive way, we have trained the LD-XGBoost in a grid-wise manner using 6-month order transaction data for the 50 restaurant-household grids within the original service network of the example city. The full transaction data covers orders placed between 10:45 a.m. to 9:30 p.m. every day, corresponding to a total number of 43 different 15-minute time windows per day. The predictors are trained to predict the number of orders placed within each 15-minute time window, using the updated lagged features summarized from the actual historical demand as inputs. For evaluation purposes, we analyze the prediction accuracy of the trained LD-XGBoost predictors over a one-month period for all restaurant-household grids. 

\begin{figure}
    \centering
    \includegraphics[width= 0.70 \textwidth]{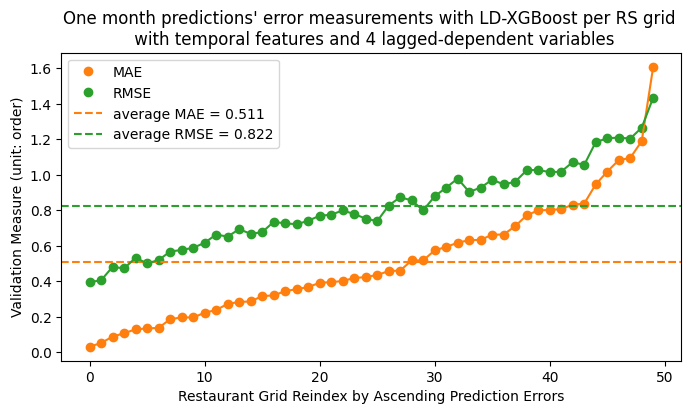}
    \caption{One-month average prediction error of per 15-minute order arrivals for 50 restaurant-household grids in the example city.}
    \label{fig:demand_accuracy}
\end{figure}

For the prediction performance metrics, we adopt the average mean absolute error (MAE) and average root-mean-squared error (RMSE), where the MAE and RMSE are measured for the forecasting accuracy of each grid, and we summarize the overall accuracy by taking the average among grids. 
The average MAE is 0.511 orders with a standard deviation of 0.335, and the average RMSE is 0.822 orders with a standard deviation of 0.239 for predicting the number of order arrivals per 15-minute time window per grid. Considering the average absolute error is less than one order, and the predictions can be made in real-time, we believe LD-XGBoost is a suitable forecasting method for generating demand prediction to assist further strategic operations. The grid-wise MAEs and RMSEs are plotted in ascending order in Figure \ref{fig:demand_accuracy}.

